\documentclass[lettersize,journal]{IEEEtran}

\usepackage{listings}
\usepackage{bm}
\usepackage{xspace}
\usepackage{verbatim}
\usepackage{tcolorbox}
\usepackage{pifont}
\usepackage{stfloats}
\usepackage{textcomp}
\usepackage{booktabs, tabularx}
\usepackage{rotating}
\usepackage{color, colortbl}
\usepackage{xcolor}
\usepackage{cite}
\usepackage{soul}
\usepackage{fancybox}
\usepackage{multirow}
\usepackage{enumitem}
\usepackage{mdframed}
\usepackage{subcaption}  
\usepackage{balance}
\usepackage[normalem]{ulem}

\usepackage{url}

\usepackage[caption=false,font=footnotesize]{subfig}
\usepackage{balance}
\usepackage{footnote}

\usepackage{array}

\usepackage{adjustbox}
\usepackage{threeparttable}

\definecolor{dkgreen}{rgb}{0,0.6,0}
\definecolor{gray}{rgb}{0.5,0.5,0.5}
\definecolor{mauve}{rgb}{0.58,0,0.82}
\definecolor{dkred}{rgb}{0.6, 0.1, 0}
\definecolor{dkblue}{rgb}{0, 0, 0.6}

\newcommand{\phead}[1]{\vspace{1mm} \noindent {\bf #1}}

\newcommand{\pattern}[1]{\emph{#1}}

\newcommand{\model}{\textsf{LLMVulExp}\xspace}

\newcommand{\greybox}[2]{\begin{tcolorbox}[width=\linewidth,boxrule=0pt,top=1pt, bottom=1pt, left=1pt,right=1pt, colback=gray!20,colframe=gray!20]
{\textbf{#1}: }{#2}
\end{tcolorbox}}

\newcommand{\revised}[1]{#1}
\newcommand{\revise}[1]{#1}

\usepackage{array}

\definecolor{mygreen}{rgb}{0,0.6,0}
\definecolor{mygray}{RGB}{245,245,244}
\definecolor{mymauve}{rgb}{0.58,0,0.82}

\mdfdefinestyle{mystyle}{
  backgroundcolor=mygray,
  leftmargin=0.45em,
  rightmargin=0.45em,
  innertopmargin=10pt,
  innerbottommargin=10pt,
  skipabove=10pt,
  skipbelow=10pt
}

\begin{document}

\title{Towards Explainable Vulnerability Detection with Large Language Models}

\author{Qiheng Mao$^{*}$, Zhenhao Li$^{*}$, 
Xing Hu, Kui Liu, Xin Xia$^{\dagger}$ and Jianling Sun

\thanks{Qiheng Mao, Xing Hu, Kui Liu, Xin Xia and Jianling Sun are with the Zhejiang University, Hangzhou, Zhejiang, China. E-mail: maoqiheng@zju.edu.cn, xinghu@zju.edu.cn, brucekuiliu@gmail.com, xin.xia@acm.org and sunjl@zju.edu.cn. Zhenhao Li is with the York University, Toronto, Canada. E-mail: Zhenhao.li@ieee.org.}
\thanks{Qiheng Mao and Zhenhao Li are co-first authors with equal contributions. Xin Xia is the corresponding author.}
}

\maketitle 

\begin{abstract}
Software vulnerabilities pose significant risks to the security and integrity of software systems. Although prior studies have explored vulnerability detection using deep learning and pre-trained models, these approaches often fail to provide the detailed explanations necessary for developers to understand and remediate vulnerabilities effectively.
The advent of large language models (LLMs) has introduced transformative potential due to their advanced generative capabilities and ability to comprehend complex contexts, offering new possibilities for addressing these challenges.
In this paper, we propose \model, an automated framework designed to specialize LLMs for the dual tasks of vulnerability detection and explanation. To address the challenges of acquiring high-quality annotated data and injecting domain-specific knowledge, \model leverages prompt-based techniques for annotating vulnerability explanations and fine-tunes LLMs using instruction tuning with Low-Rank Adaptation (LoRA), enabling \model to detect vulnerability types in code while generating detailed explanations, including the cause, location, and repair suggestions. Additionally, we employ a Chain-of-Thought (CoT) based key code extraction strategy to focus LLMs on analyzing vulnerability-prone code, further enhancing detection accuracy and explanatory depth.
We conducted experiments across multiple vulnerability detection settings on three benchmark datasets, demonstrating the effectiveness of our method.
This study highlights the feasibility of utilizing LLMs for real-world vulnerability detection and explanation tasks, providing critical insights into their adaptation and application in software security.
\end{abstract}


\section{Introduction}
\label{sec:intro}

\IEEEPARstart{S}{oftware} vulnerabilities are flaws or weaknesses in a system that can be exploited by an attacker to perform unauthorized actions~\cite{zhou2019devign,li2021vulnerability,pan2024towards}, potentially leading to severe consequences such as data breaches, financial losses, and reputational damage. The growing complexity and interconnectedness of modern software systems pose significant challenges to the timely and effective identification and mitigation of these vulnerabilities.

\begin{figure}
\centering
\includegraphics[width=\linewidth]{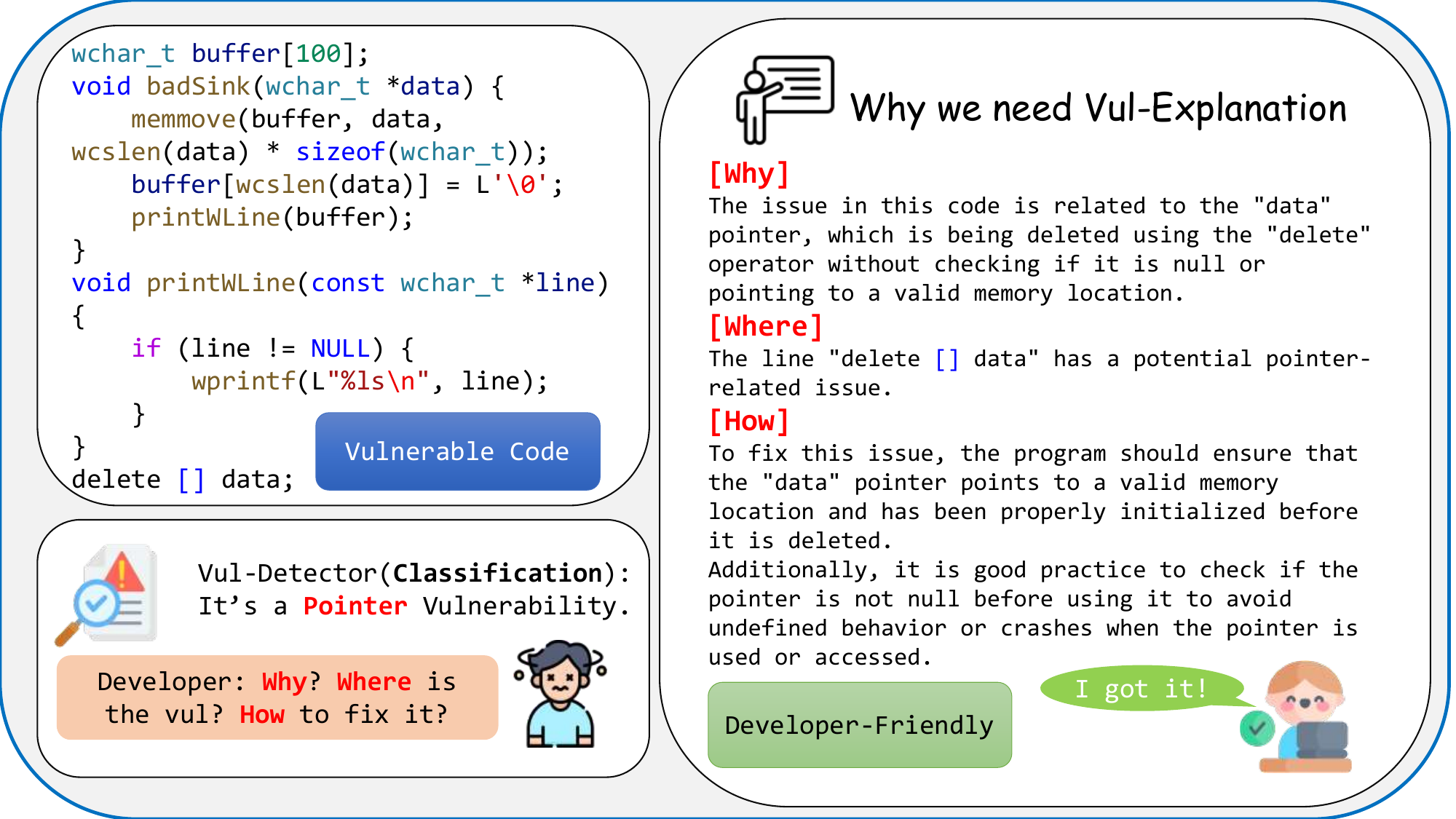}
\caption{Illustration of Vulnerability Detection with and without Explanation. An explanation provides comprehensive information that aids in understanding and fixing the vulnerability.}
\label{fig:example}
\end{figure}

Vulnerability detection techniques are predominantly divided into pattern-based and deep learning-based approaches. Pattern-based methods~\cite{yamaguchi2017pattern,rough_auditing} rely on manually crafted rules, while deep learning-based approaches~\cite{zhou2019devign,li2021vulnerability,dam2017automatic,li2018vuldeepecker,russell2018automated,qiu2024vulnerability,zhang2023learning} leverage existing vulnerability datasets and various code representation techniques to train predictive models. Despite their advancements, a critical limitation of existing methods is their failure to provide detailed, human-understandable explanations for the vulnerabilities they detect. This explanatory gap severely impedes a developer's ability to comprehend, validate, and effectively mitigate the identified risks in practice. Therefore, there is a pressing need for techniques that not only detect vulnerabilities but also provide insightful explanations, as illustrated in Figure~\ref{fig:example}. The figure contrasts a standard detection result with an enriched one, highlighting how comprehensive explanations aid in understanding and remediation.

The advent of Large Language Models (LLMs)~\cite{touvron2023llama,fu2023chatgpt,zhou2024large,zhang2024dual,yang2025code} offers a promising approach to bridging this explanatory gap. A prominent strategy to adapt these general-purpose models for specialized tasks is the ``teacher-student'' paradigm~\cite{zhou2023lima}, where a ``student'' model is fine-tuned on data generated by a powerful ``teacher'' to distill its capabilities. However, applying this paradigm to the security-critical domain of vulnerability analysis is uniquely challenging. The goal is not merely to generate descriptive text, but to instill a model with the capacity for specialized, autonomous reasoning to both detect and explain vulnerabilities in real-world code. This requires a deep understanding of code semantics and security principles—a level of expertise far beyond general Q\&A—presenting a formidable research challenge in transforming LLMs into specialized security assistants, which this work aims to address.

The primary research challenges in this endeavor are twofold:
(1) \textit{\textbf{Acquiring High-Quality Vulnerability Explanation Data at Scale:}} In the field of vulnerability detection, there is a significant lack of annotated explanation data. Additionally, annotating such data typically requires costly manual labeling by domain experts, making the collection of large-scale annotated datasets a major challenge for model fine-tuning.
(2) \textit{\textbf{Instilling Autonomous End-to-End Vulnerability Explanation Capability:}} The goal is to develop a ``student'' LLM that can autonomously analyze unfamiliar code, detect potential vulnerabilities, and generate accurate, actionable explanations—a task far more demanding than merely explaining pre-identified issues~\cite{cheng2024vulnerability}. Effectively fine-tuning a model for this integrated detection-explanation workflow, and determining if learning to explain concurrently enhances detection, remain key open questions.

To address these challenges, we introduce \model, a framework centered on a meticulously designed ``teacher-student'' paradigm that distinguishes between two LLM roles. First, we task a powerful ``teacher'' LLM with an ``open-book'' style problem: given the ground-truth vulnerability type and location, it generates a high-quality explanation articulating why the known flaw exists. This rich explanatory data is then used to train a specialized ``student'' LLM for the ``closed-book,'' real-world task of autonomously detecting and explaining vulnerabilities in unfamiliar code. Critically, this means the capability distilled from the teacher is not a magical detection ability, but rather the ability to structure and articulate complex security logic, from which the student develops its own autonomous reasoning and detection skills.

To address the first challenge, our approach leverages powerful ``teacher'' LLMs. By providing them with ground-truth vulnerability type information, we guide them to generate high-quality explanatory data at scale on three vulnerability detection benchmark: SeVC~\cite{li2021sysevr}, Diversevul~\cite{chen2023diversevul} and PrimeVul~\cite{ding2024vulnerability}. To ensure rigorous assessment, we also introduce a novel evaluation protocol for the generated explanations, comprising specific metrics—Accuracy, Clarity, and Actionability—and an automated ``LLM-as-a-judge'' methodology~\cite{cheshkov2023evaluation, li2024generation}, addressing a critical gap in evaluation standards.
To address the second challenge, we employ a sophisticated fine-tuning process. Using the synthesized data, we specialize the ``student'' LLM for its autonomous, dual-capability role: first detecting vulnerabilities in unseen code and then explaining them. Our methodology explicitly investigates whether this explanation-focused training also enhances detection accuracy. We achieve this specialization efficiently using instruction tuning~\cite{zhang2023instruction} with Low-Rank Adaptation (LoRA)~\cite{hu2021lora} and promote deeper reasoning during inference via a Chain-of-Thought (CoT) prompting strategy~\cite{CoT}.

Overall, the primary contributions of this paper are threefold\footnote{The replication package of our paper is available~\cite{replication_package}.}:

\begin{itemize}[leftmargin=1.5em]
\item We introduce \model, a novel ``teacher-student'' framework whose core technical innovation is a structured pipeline that leverages guided LLMs to resolve the critical data bottleneck in training explainable vulnerability detection models.
\item We propose and validate a new evaluation protocol for vulnerability explanations, featuring dedicated metrics (Accuracy, Clarity, Actionability) and an ``LLM-as-a-judge'' method, thereby advancing the assessment standards for LLM-generated security analyses.
\item Through extensive experiments, we demonstrate that our fine-tuning process equips ``student'' LLMs (e.g., CodeLlama) to autonomously detect vulnerabilities and generate accurate, clear, and actionable explanations. Our results provide valuable insights into the practical application of specialized models for end-to-end vulnerability detection and explanation.
\end{itemize}

\phead{Paper Organization.}
Section~\ref{sec:related} summarizes the related work.
Section~\ref{sec:methodology} presents the methodology of our study.
Section~\ref{sec:results} discusses the results of our research questions.
Section~\ref{sec:discussion} discusses the implications of our study.
Section~\ref{sec:threats} discusses the threats to validity.
Section~\ref{sec:conclusion} concludes the paper.
\section{Related Work}
\label{sec:related}

In this section, we review the related work in two key areas: vulnerability detection and explanation.

\subsection{Vulnerability Detection}
Prior studies proposed a series of deep learning approaches~\cite{zhou2019devign,li2021vulnerability,dam2017automatic,li2018vuldeepecker,russell2018automated,tan2024similar,liu2025industrial,zhang2023vulnerability} to detect vulnerabilities. These methods have employed labeled vulnerability data to train neural networks, enabling the models to capture semantic features associated with vulnerabilities.

The advent of LLMs has significantly influenced the field of vulnerability detection. Techniques such as zero-shot prompting~\cite{zhou2024large, fu2023chatgpt}, in-context learning~\cite{dong2022survey, brown2020language}, and fine-tuning~\cite{hu2021lora,xu2023parameter, chen2023diversevul} have been explored to enhance LLM-based vulnerability detection. 
Cheshkov et al.\cite{cheshkov2023evaluation} evaluated ChatGPT and GPT-3, highlighting their inability to classify vulnerable code accurately in binary and multi-label settings. 
Gao et al.\cite{gao2023far} proposed a benchmark for LLMs in vulnerability detection, demonstrating that with few-shot prompting on simpler datasets, LLMs can perform comparably to deep learning-based methods. 
Nong et al.\cite{nong2024chain} showed that chain-of-thought prompting, based on the semantic structure of code, improves detection accuracy. 
Sun et al.\cite{sun2024llm4vuln} found that supplementing LLMs with high-quality vulnerability-related knowledge enhances their performance. 
Yusuf et al.\cite{yusuf2024your} observed that natural language instructions boost vulnerability detection across multiple programming languages. 
Steenhoek et al.\cite{steenhoek2024comprehensive} surveyed eleven LLMs, indicating the limitations of applying LLMs directly to vulnerability detection without fine-tuning.

These studies underscore the challenges and potential of using LLMs for vulnerability detection. 
Unlike these methods, our research focuses on more practical and challenging explanation tasks, and we enhance the vulnerability understanding and analysis capabilities of LLMs through specialized fine-tuning.

\subsection{Vulnerability Explanation}

Vulnerability explanation in software security involves providing human-understandable insights into why a piece of code is vulnerable—encompassing its cause, potential exploitation, impact, and remediation—thereby extending beyond mere statement-level localization. Despite its critical importance for aiding developers and reducing reliance on security experts, achieving comprehensive explanations with deep learning models remains a significant, underexplored challenge.

Prior works have made strides in aspects related to Vulnerability explanation. Many, such as VulDeeLocator~\cite{li2021vuldeelocator}, IVDetect~\cite{li2021vulnerability}, LineVul~\cite{fu2022linevul}, VELVET~\cite{ding2022velvet}, and LineVD~\cite{hin2022linevd}, focus on statement-level vulnerability localization, pinpointing vulnerable code lines using techniques like Bi-LSTMs or graph-based analysis. While this localization is a crucial first step, the outputs often require considerable human interpretation to form a complete explanation. Other studies like VulTeller~\cite{zhang2023learning} and VulExplainer~\cite{cheng2024vulnerability} offer deeper structural insights that contribute to detection. However, these approaches generally do not generate the complete, natural-language narratives necessary to fully explain the vulnerability's nature and implications to developers.

Accurately generating comprehensive and human-readable explanations remains a challenge for existing neural models. Our work explores fine-tuning specialized Large Language Models for both vulnerability detection and, crucially, for producing richer, more holistic explanations. We posit that LLMs, with their strong semantic understanding and text generation capabilities, can synthesize information to elucidate the 'why' and 'how' of vulnerabilities in coherent language. This aims to move beyond localization or basic structural cues, addressing a critical gap by providing actionable, understandable feedback directly to developers.

\section{Methodology}
\label{sec:methodology}

\begin{figure*}
 \centering
\includegraphics[width=0.9\linewidth ]{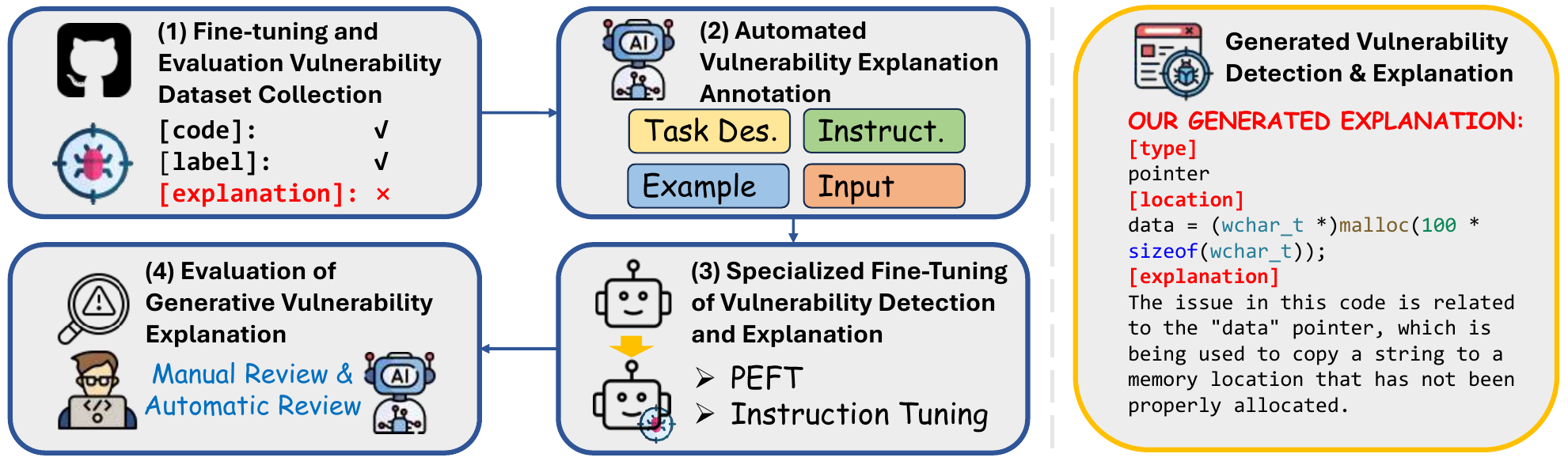}
 \caption{Overview of the \model Framework for Explainable Vulnerability Detection via LLM Fine-Tuning.}
 \label{fig:overall}
 \end{figure*}

To address the current gap in generative vulnerability explanation models and enhance the ability of LLMs to detect and explain software vulnerabilities, we propose a comprehensive framework for fine-tuning and evaluating specialized LLMs for both vulnerability detection and explanation tasks. 
Figure~\ref{fig:overall} presents an overview of our framework, namely, \model.
Specifically, our framework consists of four core stages: \ding{182} fine-tuning and evaluation vulnerability dataset collection, \ding{183} automated vulnerability explanation annotation based on prompt engineering, \ding{184} specialized fine-tuning of vulnerability detection and explanation through instruction-based fine-tuning, and \ding{185} evaluation of generative vulnerability explanation capabilities.

\phead{\ding{182} Fine-tuning and Evaluation Vulnerability Dataset Collection:} 
Enhancing the specialized capabilities of LLMs requires large quantities of high-quality domain-specific data. In the context of vulnerability detection and explanation, it is crucial to ensure the authenticity of the vulnerability code, the diversity of vulnerability types, and the sufficiency of examples for each type. In this paper, we conduct the study on three C/C++ function-level vulnerability datasets: 
(1) SeVC~\cite{li2021sysevr}, which contains four core vulnerability types and over 50,000 vulnerable code snippets,
(2) DiverseVul~\cite{chen2023diversevul}, which covers 295 real open-source projects and 150 CWE types and 
(3) PrimeVul~\cite{ding2024vulnerability}, a new vulnerability dataset with high-quality, accurately labeled data and stringent
de-duplication to offer realistic training and testing data for vulnerability detection.
We select C/C++ as the programming language due to its diverse range of vulnerability types and examples, as well as the substantial dependence of fine-tuning on the availability of large-scale data.

\begin{table}
    \caption{Statistics of the studied datasets.}
    \centering
    \renewcommand{\arraystretch}{1.2}
    \resizebox{\linewidth}{!} {
    \begin{tabular}{l|ccccc}
        \hline
\rowcolor[HTML]{EFEFEF}  \textbf{Dataset}  &  \textbf{Non-Vul \#} & \textbf{Ori. Vul \#}   &\textbf{Ann. Vul \#} &\textbf{Vul-Type \#}  &\textbf{Eval. Setting} \\
        \hline
         \textbf{SeVC} &  364,232 & 56,395   & 40,491 & 4 & Single/Multi-Type\\ \hline
        \textbf{DiverseVul}   & 330,492 & 18,945   & 9,161 & 10 & Multi-Label\\  \hline
        \textbf{PrimeVul}   & 228,800 & 6,968 & 4,041 & 10 & Multi-Class\\ \hline
    \end{tabular}
    }
    \label{table:rq1_dataset}
\end{table}

\noindent \textit{\underline{Semantics-based Vulnerability Candidate (SeVC)}} dataset includes 126 distinct Common Weakness Enumeration (CWE) types, comprising 56,395 vulnerable samples and 364,232 non-vulnerable ones.
The SeVC dataset is categorized into four primary vulnerability classes based on the underlying causes: \pattern{Library/API Function Call}, \pattern{Array Usage}, \pattern{Pointer Usage} and \pattern{Arithmetic Expression}. 
Each category contains a substantial number of samples, making \pattern{SeVC} ideal for fine-tuning and evaluating models designed to detect and explain specific vulnerabilities. Therefore, we conduct both binary and multi-class detection tasks on the SeVC dataset.

\noindent \textit{\underline{DiverseVul}} is a comprehensive C/C++ vulnerability dataset that includes 18,945 vulnerable functions and 330,492 non-vulnerable functions, extracted from 7,514 commits, covering 150 CWEs. As the largest real-world C/C++ vulnerability dataset, DiverseVul is characterized by longer code snippets, a broader range of projects, more diverse vulnerability types, and lower label noise. It serves as a benchmark for evaluating the capability of LLMs in handling real-world scenarios with a wider variety of vulnerabilities. For our fine-tuning and evaluation target, we select the top ten most frequent CWE types, ensuring a sufficient number of samples and a well-defined number of classes for our multi-class detection task.

\noindent \textit{\underline{PrimeVul}} is a state-of-the-art benchmark for C/C++ vulnerability detection, designed to overcome the critical data quality issues of prior datasets. It ensures high label accuracy (up to 92\% ) through novel automated techniques, such as correlating code changes directly with official NVD descriptions and analyzing single-function security commits. This high-fidelity labeling, combined with stringent de-duplication and realistic chronological data splitting, presents a uniquely challenging testbed. Its difficulty is underscored by the original study, which shows that the performance of powerful models plummets on this benchmark—F1 scores drop from over 68\% on older datasets to just 3.09\%, and even GPT-4 fails to outperform random guessing on its challenging paired evaluation. Therefore, PrimeVul serves as a stringent benchmark to assess a model's ability to generalize and reason about complex, real-world vulnerabilities, making it an ideal choice for validating our approach.

We first deduplicate all samples using the SHA-256 hash method and then partition the data into training (80\%), validation (10\%), and test (10\%) sets. To save training cost and mitigate model bias arising from severe class imbalance, we create balanced training and validation sets by downsampling non-vulnerable samples to a 1:1 ratio. The handling of the test set is tailored to the dataset's characteristics. For the two datasets composed of real-world vulnerabilities, we preserve the original, imbalanced class distribution to ensure the evaluation reflects a realistic deployment scenario. In contrast, for the SeVC dataset, which contains a larger number of synthetic vulnerability samples, we use a balanced test set to facilitate a more effective multi-class evaluation. This overall approach allows for unbiased model training while rigorously testing its practical effectiveness. The final dataset statistics are detailed in Table~\ref{table:rq1_dataset}.

\phead{\ding{183} Automated Vulnerability Explanation Annotation:} Traditional open-source vulnerability datasets mainly contain source code, vulnerability labels, CWE types, and commit messages. However, they often lack detailed explanations of the vulnerability logic within the source code, posing a significant challenge for vulnerability detection techniques that aim to provide meaningful explanations for detected issues. Manually annotating real-world vulnerable code explanations requires extensive software development experience and a deep understanding of software vulnerabilities, which incurs high labor and time costs~\cite{lu2023llama}.
As the scale of manually annotated data remains a significant challenge, an increasing number of researchers are leveraging the powerful generative capabilities of LLMs to synthesize data. This approach has been validated in various domains through improvements in downstream task performance metrics. However, in the specialized field of vulnerability detection, where domain expertise is critical, the feasibility of synthesizing data has yet to be fully validated.

To address this challenge, we introduce an innovative automated vulnerability explanation annotation method based on prompt engineering with LLMs. This method capitalizes on the contextual learning and instruction-following capabilities of LLMs, enabling large-scale, high-quality automated synthesis of vulnerability explanations.
Our approach decomposes the explanation task into three sub-goals: (1) vulnerability discrimination, (2) identifying the location of the vulnerability in the code, and (3) providing a specific explanation. The model is guided by instruction-based prompt templates, paired with well-annotated examples, which stimulate its contextual learning capabilities, ensuring that the generated explanations are both accurate and informative.

We implement this annotation process using LLM, selecting appropriate models based on a balance of cost, efficiency, and capability. For our initial annotation of the SeVC and DiverseVul datasets, we employed GPT-3.5~\cite{gpt3}, accessed via the API provided by OpenAI~\cite{openai_api}. For the subsequently added experiments and the PrimeVul dataset, we utilized Deepseek-V3~\cite{liu2024deepseek}, a state-of-the-art model noted for its strong code processing capabilities.
This dual-model approach allowed us to complete a large-scale annotation effort, resulting in 40,491, 9,161 and 4,041 vulnerability explanation data points across the respective datasets.
This effort fills a significant gap in the availability of vulnerability explanation annotations, enabling more effective training and evaluation of vulnerability detection and explanation models.
To validate the effectiveness of the synthesized data, we indirectly evaluated the performance of the fine-tuned model in vulnerability detection tasks across multiple scenarios. Additionally, we propose evaluation metrics for vulnerability explanation quality, along with both manual and automated evaluation methods, to directly assess the quality of the generated explanations.

\phead{\ding{184} Specialized Fine-Tuning of Vulnerability Detection and Explanation:} The automated annotation of vulnerability explanations for open-source data creates a large-scale dataset, effectively addressing data bottlenecks in the fine-tuning process for vulnerability detection and explanation. 
To enhance the detection and explanation capabilities of LLMs (especially open-source models with lower computational overhead), we fine-tune general LLMs to specialize in detecting and explaining specific types of vulnerabilities in real-world code. Instruction-based prompts are used to guide these tasks, helping LLMs to accurately understand the objectives and generate standardized outputs. To minimize the computational cost of the fine-tuning process, we employ the parameter-efficient fine-tuning method LoRA~\cite{hu2021lora}, which significantly reduces both time and space requirements.

\phead{\ding{185} Evaluation of Generative Vulnerability Explanation:} 
The limited research on model-generated vulnerability explanations underscores the need for robust evaluation methods tailored to LLM-generated outputs. Similar to annotation challenges, manual evaluation requires considerable human and time resources. Moreover, effective vulnerability explanations must be assessed across multiple dimensions to ensure they offer practical value to developers in real-world scenarios.
To address this, we propose an evaluation framework based on three critical dimensions: accuracy, clarity, and actionability. These dimensions collectively measure the correctness, comprehensibility, and practical applicability of the generated explanations. To enhance evaluation efficiency, we introduce an automated evaluation method based on LLMs. This method leverages prompt engineering to enhance evaluation efficiency while ensuring reliable assessments. Additionally, expert manual verification is employed to validate both the quality of the outputs from the specialized vulnerability explanation model and the feasibility of the LLM-based automated evaluation scheme.

\subsection{Vulnerability Interpretation Enhancement Prompting}
Despite the robust code understanding and analysis capabilities of LLMs, they face significant challenges in complex reasoning tasks that demand a deep understanding of code, advanced reasoning abilities, and specialized knowledge of vulnerabilities. These challenges often result in insufficient detection accuracy and vague vulnerability analyses. To address these limitations, we integrate instruction-based fine-tuning techniques with the contextual learning capabilities of LLMs. By leveraging prior knowledge from open-source code vulnerabilities, we design prompt templates for data annotation, fine-tuning, reasoning, and evaluation, effectively enabling the application of LLMs in vulnerability detection and explanation tasks across all critical stages of the framework.

\begin{figure}
 \centering
\includegraphics[width=1.00\linewidth ]{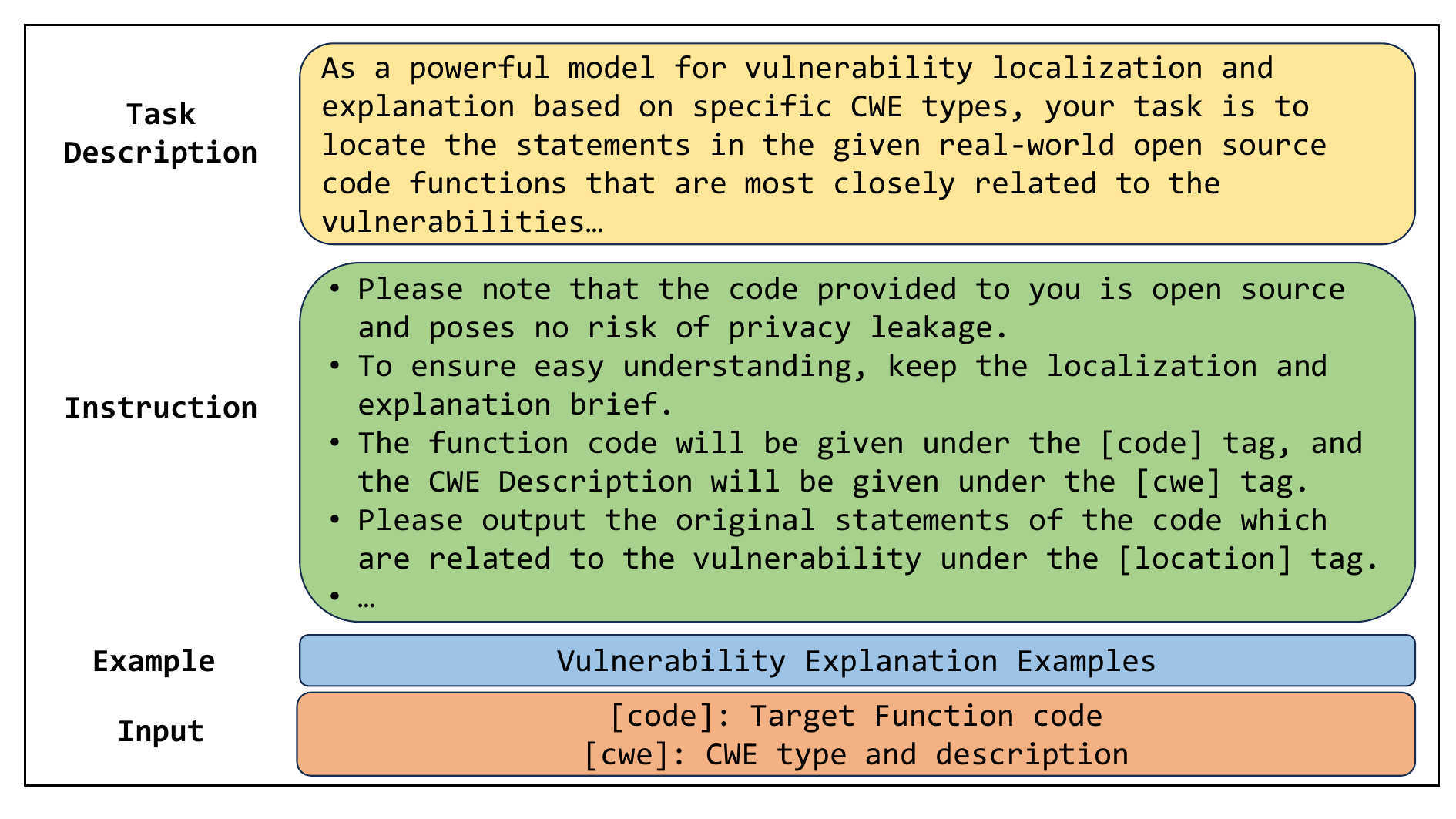}
 \caption{Prompt template of Explanation Annotation.}
 \label{fig:prompt-label}
 \end{figure}

Figure~\ref{fig:prompt-label} illustrates the structure of our prompt templates. These templates are composed of four main components: task description, specific instructions, generation examples, and sample input. \textbf{(1) Task description: } Specifies the template for the current vulnerability detection and explanation task, including details about the types of vulnerabilities being addressed and the basic input-output format. This component provides the LLM with a clear understanding of task requirements and relevant background knowledge.
\textbf{(2) Specific instructions: } Define the required input-output format for the LLM, such as step-by-step output guidelines, the range of vulnerabilities to focus on, and output length constraints. These instructions utilize the LLM's instruction-following ability to ensure standardized and uniform outputs, which facilitates subsequent processing and analysis.
\textbf{(3) Generation examples: } Contain manually curated samples of vulnerable code snippets paired with corresponding explanation data. These examples help the LLM better comprehend the task's goals and improve the quality of generated outputs.
\textbf{(4) Sample input: } Includes the code to be analyzed and, during the annotation phase, incorporates associated labels and supplementary information such as CWE types, CVE descriptions, or commit messages.

Through rigorous testing and evaluation across various vulnerability detection and explanation tasks, we validate that these prompt templates effectively achieve their intended objectives at all stages. The resulting fine-tuned models demonstrate improved performance in specialized vulnerability detection and explanation tasks.

\subsection{Key Code Extraction Based on Chain-of-Thought (CoT)}
One of the key challenges in explaining code vulnerabilities lies in accurately identifying and extracting critical code segments—referred to as key code—from lengthy code snippets. Key code typically includes statements, variable structures, and other components that are closely related to the vulnerability type under investigation. These elements often represent the root cause or propagation paths of vulnerabilities and provide essential context for understanding the issue. However, traditional static analysis techniques, such as abstract syntax trees, program dependence graphs, and control flow graphs, often fail to generalize effectively in diverse and complex vulnerability detection scenarios.

\begin{figure}
 \centering
\includegraphics[width=1.00\linewidth ]{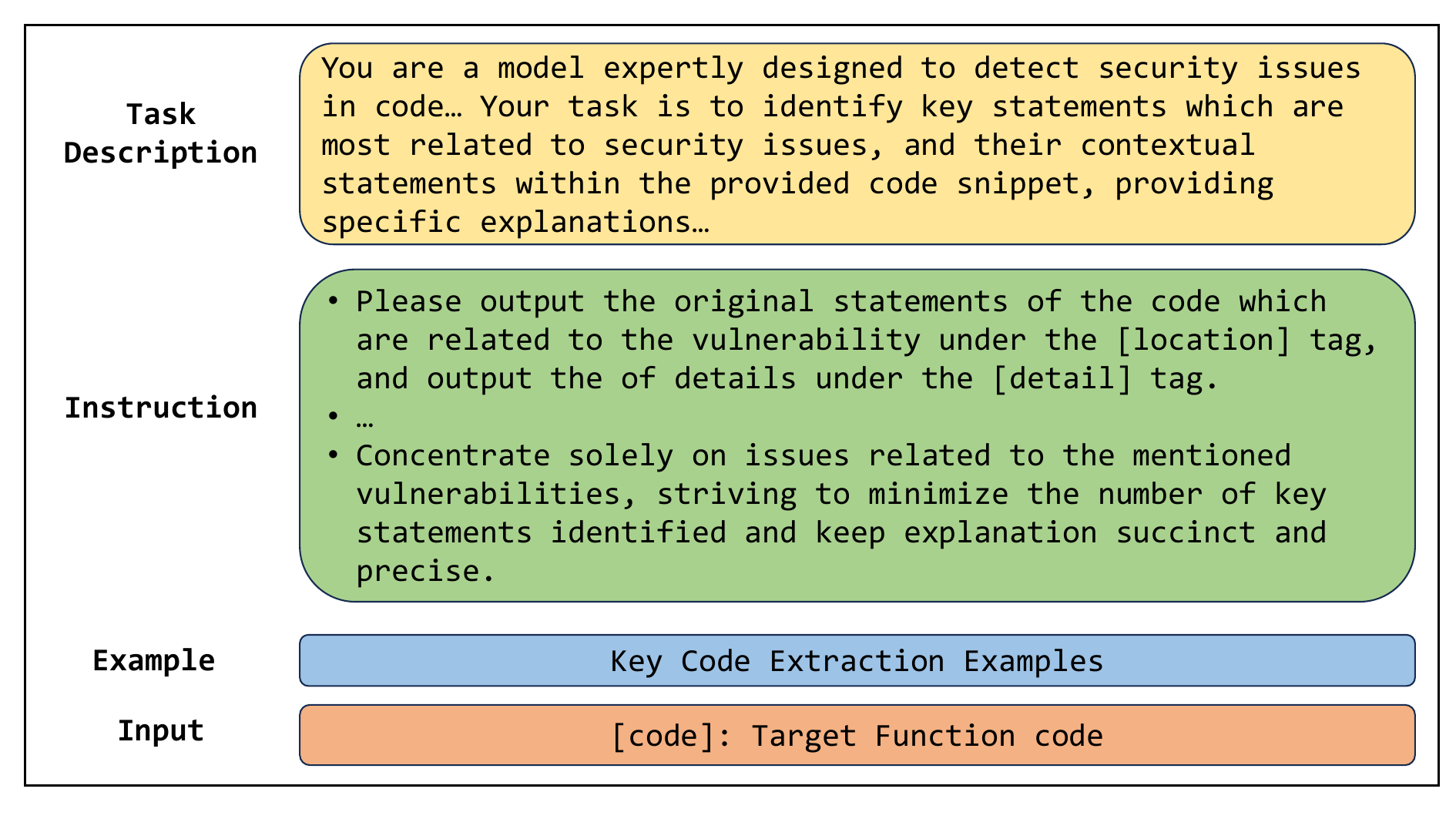}
 \caption{Prompt template of Key Code Extraction.}
 \label{fig:prompt-key}
 \end{figure}

To address these challenges and improve the capability of large models to analyze complex code structures, we propose a novel Chain-of-Thought (CoT) enhancement method that incorporates automated key code extraction. This approach enables the model to focus on the most relevant parts of the code, thereby improving accuracy and interpretability.
Using the prompt template illustrated in Figure~\ref{fig:prompt-key}, we employ LLMs to automatically extract key statements by analyzing the semantic context of the code and the specific vulnerability type. These extracted statements form a structured CoT, guiding the fine-tuned model through a step-by-step process for targeted vulnerability detection, location, and explanation. For example, in a buffer overflow scenario, key code may include array declarations, bounds-checking statements, or pointer dereferences that contribute directly to the vulnerability. By integrating these extracted key statements into the prompts during fine-tuning, the model can systematically reference this crucial information, enhancing both the interpretability and precision of the vulnerability explanation process.

This CoT-based key code extraction methodology ensures that the model is not only detecting vulnerabilities more accurately but also providing detailed, context-rich explanations that developers can use to understand and address these issues effectively.

\section{Results}
\label{sec:results}
In this section, we discuss the results by proposing and answering the following research questions:

\begin{itemize}[leftmargin=*]
    \item \textbf{RQ1}: How effective are LLMs in detecting software vulnerabilities?
    
    \item \textbf{RQ2}: How proficient are LLMs in explaining the detected vulnerabilities?

    \item \textbf{RQ3}: How do explanations affect the results of vulnerability detection?

    \item \textbf{RQ4}: How does the key code extraction impact detection and explanation performance?
\end{itemize}

\subsection{\textbf{RQ1: How effective are LLMs in detecting software vulnerabilities?}} \label{sec:RQ1}
Accurate detection is the foundation for any subsequent vulnerability analysis or explanation. In this RQ, we first conduct a comprehensive evaluation of our fine-tuned model's detection performance against strong baselines across two distinct, real-world scenarios.

\subsubsection{\textbf{Experimental Setup.}}

\phead{Datasets and Tasks.}
As detailed in Section~\ref{sec:methodology}, we select three benchmarks for evaluation: 
\pattern{SeVC}~\cite{li2021sysevr},
\pattern{DiverseVul}~\cite{chen2023diversevul} and \pattern{PrimeVul}~\cite{ding2024vulnerability}. Based on their distinct characteristics, we formulate corresponding detection tasks:
\begin{itemize}[leftmargin=1.5em, itemsep=0pt, topsep=3pt]
    \item On \textbf{\pattern{SeVC}}, we peform binary classification vulnerability toward single-type vulnerability.
    \item On \textbf{\pattern{SeVC}}, we perform coarse-grained multi-class vulnerability detection.
    \item On \textbf{\pattern{DiverseVul}}, we perform multi-label CWE classification (a function can have multiple vulnerability types) and binary classification.
    \item On \textbf{\pattern{PrimeVul}}, we perform multi-class CWE classification (a function has exactly one vulnerability type) and binary classification.
\end{itemize}

\phead{Model and Baselines.}
Our primary model, \model, is based on \pattern{CodeLlama-13B-Instruct}~\cite{codellama_model}, fine-tuned using the LoRA technique. To rigorously situate its performance, we compare it against a comprehensive set of four baselines: two general-purpose pre-trained models, \pattern{CodeT5}~\cite{wang2021codet5} and \pattern{CodeBERT}~\cite{feng2020codebert}, and two state-of-the-art specialized vulnerability detectors, \pattern{LineVul}~\cite{fu2022linevul} and \pattern{CausalVul}~\cite{rahman2024towards}. While our ultimate goal includes evaluating explanation capabilities, validating detection efficacy against these strong baselines is a critical first step.

\phead{Metrics and Implementation Details.}
We use Precision, Recall, and F1-Score as our primary evaluation metrics. For the multi-label task on \pattern{DiverseVul}, we report Micro-F1 and Macro-F1. For the multi-class task on \pattern{SeVC} and \pattern{PrimeVul}, we report Weighted-F1 and Macro-F1 to analyze performance across its imbalanced class distribution. We implement our approach using the Transformers~\cite{thapa2022transformer} and PEFT~\cite{peft} libraries. All experiments are conducted on two NVIDIA A100-80GB GPUs. Key hyperparameters include a learning rate of 3e-4, 3 training epochs, and a batch size of 2. For the LoRA configuration, we set  \{\texttt{q\_proj}, \texttt{k\_proj}, \texttt{v\_proj}, \texttt{o\_proj}, \texttt{up\_proj}, \texttt{down\_proj}, \texttt{gate\_proj}\}. The LoRA rank is set to 16, the LoRA scaling factor to 16/32, and the dropout rate to 0.05.

\subsubsection{\textbf{RQ1.1: Detection with specified vulnerability type using a dedicated model.}} 

To evaluate the detection capability for a single type of vulnerability, we fine-tune a separate model for each vulnerability type on \pattern{SeVC}, resulting in specialized models tailored to individual vulnerability types. The sample sizes of each type is shown in Table~\ref{table:rq1.0}. In both training and inference prompts, explicitly specify the target vulnerability type to narrow the scope of detection and improve precision.  During training, the model utilizes explanations annotated by GPT-3.5, which include detailed information about the vulnerability location and its specific characteristics, as the target output for performing both detection and explanation tasks. The model's input during both training and inference consists of code snippets from the corresponding dataset examples.
As a generative model, the task is formulated as a binary classification based on the semantics of the generated text. For vulnerability samples, the model generates corresponding explanatory information, while for non-vulnerability samples, it produces a predefined fixed pattern (e.g., \textit{``There are no security issues''}).

\begin{table}[t]
    \centering
    \small
    \caption{Performances Comparison of fine-tuning on SeVC(RQ1.1).}
        \resizebox{\columnwidth}{!} {
    \tabcolsep=3pt
    \renewcommand\arraystretch{1.10}
    \begin{tabular}{lccccc}
    \hline
    Metric & API & Arith. & Pointer & Array & Average \\
    \hline
    \#Samples & 20,294 & 5,968 & 38,040 & 16,680 & 80,982 \\
    \hline
    Precision (\model) & \cellcolor[HTML]{EFEFEF}\textbf{91.6\%} & \cellcolor[HTML]{EFEFEF}\textbf{90.3\%} & \cellcolor[HTML]{EFEFEF}\textbf{93.7\%} & \cellcolor[HTML]{EFEFEF}\textbf{95.3\%} & \cellcolor[HTML]{EFEFEF}\textbf{92.7\%} \\
    Precision (CodeLlama) & 61.8\% & 64.5\% & 70.2\% & 66.2\% & 65.7\% \\
    \hline
    Recall (\model) & \cellcolor[HTML]{EFEFEF}\textbf{94.5\%} & \cellcolor[HTML]{EFEFEF}\textbf{93.6\%} & \cellcolor[HTML]{EFEFEF}\textbf{91.2\%} & \cellcolor[HTML]{EFEFEF}\textbf{89.7\%} & \cellcolor[HTML]{EFEFEF}\textbf{92.3\%} \\
    Recall (CodeLlama) & 26.3\% & 30.4\% & 50.2\% & 36.6\% & 35.9\% \\
    \hline
    F1 (\model) & \cellcolor[HTML]{EFEFEF}\textbf{93.0\%} & \cellcolor[HTML]{EFEFEF}\textbf{91.9\%} & \cellcolor[HTML]{EFEFEF}\textbf{92.4\%} & \cellcolor[HTML]{EFEFEF}\textbf{92.4\%} & \cellcolor[HTML]{EFEFEF}\textbf{92.4\%} \\
    F1 (CodeLlama) & 36.9\% & 41.4\% & 58.6\% & 47.2\% & 46.0\% \\
    \hline
    \end{tabular}
    }
    \label{table:rq1.0}
\end{table}

\begin{table*}[ht]
    \centering
    \caption{Performances of Single-Type Detection on SeVC (RQ1.1).}
    \begin{tabular}{lccccccccc}
    \hline
    & \multicolumn{3}{c}{Precision} & \multicolumn{3}{c}{Recall} & \multicolumn{3}{c}{F1} \\
    \cmidrule(lr){2-4} \cmidrule(lr){5-7} \cmidrule(lr){8-10}
    Type & Ours & CodeT5 & CodeBert & Ours & CodeT5 & CodeBert & Ours & CodeT5 & CodeBert \\
    \hline
    API & 91.6\% & 92.2\% & \cellcolor[HTML]{EFEFEF}\textbf{93.2\%} & \cellcolor[HTML]{EFEFEF}\textbf{94.5\%} & 88.1\% & 87.1\% & \cellcolor[HTML]{EFEFEF}\textbf{93.0\%} & 90.1\% & 90.0\% \\
    
    Arithmetic & 90.3\% & 88.2\% & \cellcolor[HTML]{EFEFEF}\textbf{90.9\%} & 93.6\% & 94.7\% & \cellcolor[HTML]{EFEFEF}\textbf{98.0\%} & \cellcolor[HTML]{EFEFEF}\textbf{91.9\%} & 91.3\% & \cellcolor[HTML]{EFEFEF}\textbf{91.9\%} \\
    
    Pointer & 93.7\% & 93.5\% & \cellcolor[HTML]{EFEFEF}\textbf{95.8\%} & 91.2\% & \cellcolor[HTML]{EFEFEF}\textbf{95.0\%} & 93.4\% & 92.4\% & 94.3\% & \cellcolor[HTML]{EFEFEF}\textbf{94.6\%} \\
    
    Array & \cellcolor[HTML]{EFEFEF}\textbf{95.3\%} & 92.9\% & 95.1\% & 89.7\% & 94.1\% & \cellcolor[HTML]{EFEFEF}\textbf{92.2\%} & 92.4\% & 93.5\% & \cellcolor[HTML]{EFEFEF}\textbf{93.6\%} \\
    Average & 92.7\% & 91.7\% & \cellcolor[HTML]{EFEFEF}\textbf{93.7\%} & 92.3\% & 93.0\% & \cellcolor[HTML]{EFEFEF}\textbf{92.7\%} & \cellcolor[HTML]{EFEFEF}\textbf{92.4\%} & 92.3\% & 92.2\% \\
    \hline
    \end{tabular}
    
    \label{table:rq1.1}
\end{table*}

\phead{Experimental Results.}
We first evaluate the impact of fine-tuning on the detection accuracy of CodeLlama. As shown in Table~\ref{table:rq1.0}, the original CodeLlama-13B-Instruct model lacks precise vulnerability identification capabilities, demonstrating suboptimal performance in detecting specific types of vulnerabilities. However, fine-tuning significantly improves detection accuracy. Consequently, we exclude the original CodeLlama from subsequent comparisons.

From Table~\ref{table:rq1.1}, we observe that our generative model achieves performance comparable to other fine-tuned classification models. This result highlights two key findings: (1) Our approach effectively enhances the model's understanding of the specific vulnerability type, enabling it to capture critical patterns for accurate detection. (2) The detection task for a single type of vulnerability is relatively less challenging for LLMs. In practical scenarios where the focus is on a limited set of vulnerability types, fine-tuning with target-specific data can yield superior performance.

\subsubsection{\textbf{RQ1.2: Detection with identification of multiple vulnerability types.}} 
In real software development environments, vulnerability risks are often diverse, posing challenges in accurately predicting the specific types of vulnerabilities present in the code. This necessitates a model capable of simultaneously detecting, classifying, and explaining various vulnerability types.  
To evaluate the effectiveness of such a model, we train a single model on \pattern{SeVC} for multi-class vulnerability detection using all available samples. Specifically, a `[type]' tag is appended to the model's output to classify the detected vulnerabilities, encompassing non-vulnerable code and the four distinct types of vulnerabilities present in \pattern{SeVC}. Unlike the previous section, where each model is trained to detect a single type of vulnerability, here we use a unified approach to enable one model to handle all types concurrently. 
During training and inference, the specific vulnerability type of detected code will not appear in the prompt. Instead, a task description encompassing all four vulnerability types in the current multi-class scenario is provided, enabling the model to develop the capability to distinguish between multiple types of vulnerabilities.

\begin{table*}
    \centering
    \caption{Performances of Multi-Type Detection on SeVC (RQ1.2).}
    \renewcommand{\arraystretch}{0.99} 
    \begin{tabular}{lccccccccc}
    \hline
    & \multicolumn{3}{c}{Precision} & \multicolumn{3}{c}{Recall} & \multicolumn{3}{c}{F1} \\
    \cmidrule(lr){2-4} \cmidrule(lr){5-7} \cmidrule(lr){8-10}
    Type & Ours & CodeT5 & CodeBert & Ours & CodeT5 & CodeBert & Ours & CodeT5 & CodeBert \\
    \hline
    Non-vul & \cellcolor[HTML]{EFEFEF}\textbf{95.4\%} & 94.7\% & 94.6\% & \cellcolor[HTML]{EFEFEF}\textbf{98.0\%} & 94.6\% & 96.7\% & \cellcolor[HTML]{EFEFEF}\textbf{96.7\%} & 94.7\% & 95.7\% \\
    
    Array & \cellcolor[HTML]{EFEFEF}\textbf{61.9\%} & 56.0\% & 58.4\% & 55.2\% & \cellcolor[HTML]{EFEFEF}\textbf{65.6\%} & 62.8\% & 58.3\% & \cellcolor[HTML]{EFEFEF}\textbf{60.4\%} & 60.5\% \\
    
    Pointer & 72.9\% & 72.2\% & \cellcolor[HTML]{EFEFEF}\textbf{74.6\%} & 70.9\% & 66.8\% & \cellcolor[HTML]{EFEFEF}\textbf{70.9\%} & 71.9\% & 69.4\% & \cellcolor[HTML]{EFEFEF}\textbf{70.0\%} \\
    
    API & \cellcolor[HTML]{EFEFEF}\textbf{48.8\%} & 45.4\% & 44.5\% & \cellcolor[HTML]{EFEFEF}\textbf{46.0\%} & 40.8\% & 43.5\% & \cellcolor[HTML]{EFEFEF}\textbf{47.4\%} & 43.0\% & 44.0\% \\
    
    Arithmetic & \cellcolor[HTML]{EFEFEF}\textbf{70.7\%} & 65.3\% & 68.3\% & \cellcolor[HTML]{EFEFEF}\textbf{91.6\%} & 88.6\% & 87.3\% & \cellcolor[HTML]{EFEFEF}\textbf{79.8\%} & 75.2\% & 76.7\% \\
    \hline
    Weighted & \cellcolor[HTML]{EFEFEF}\textbf{79.9\%} & 78.2\% & 78.9\% & \cellcolor[HTML]{EFEFEF}\textbf{80.5\%} & 78.1\% & 79.1\% & \cellcolor[HTML]{EFEFEF}\textbf{80.1\%} & 78.0\% & 79.1\% \\
    \hline
    Macro & \cellcolor[HTML]{EFEFEF}\textbf{70.0\%} & 66.7\% & 68.1\% & \cellcolor[HTML]{EFEFEF}\textbf{72.4\%} & 71.3\% & 71.3\% & \cellcolor[HTML]{EFEFEF}\textbf{70.8\%} & 68.5\% & 69.4\% \\
    \hline
    \end{tabular}
    
    \label{table:rq1.3}
\end{table*}

\phead{Experimental Results.}
The metrics for each type in the multi-class vulnerability detection task are presented in Table~\ref{table:rq1.3}. Based on the experimental results, we observe the following:
(1)The non-vulnerable code category exhibits high precision and recall, indicating that the model maintains strong capability in identifying non-vulnerable code even in a multi-type scenario.
(2) Compared to detection results where the type of interest is provided, the overall performance shows varying degrees of decline, indicating that the model faces greater difficulty in distinguishing between vulnerability types.
(3) Among the vulnerability types, Arithmetic Expression achieves relatively better performance, whereas the remaining three types demonstrate significant confusion, making them harder to identify.
These findings indicate that, while the model is effective at recognizing non-vulnerable code, its ability to differentiate between multiple vulnerability types is limited, particularly when certain types are underrepresented in the training data.
This highlights the critical role of balanced and representative training datasets in improving the model’s performance across all vulnerability types.

\subsubsection{\textbf{RQ1.3: Performance on the DiverseVul Dataset}}

\phead{Task Setup.}
To evaluate our model's effectiveness in a standard, realistic scenario, we utilize the \pattern{DiverseVul} dataset. This benchmark is particularly suitable as it reflects real-world conditions where code can contain multiple, distinct types of vulnerabilities. We assess the models on two fronts: (1) a multi-label classification task on the top 10 most frequent CWE types, and (2) a standard binary classification task.

\phead{Experimental Results.}
The comprehensive results on \pattern{DiverseVul} are presented in Table~\ref{table:rq1.3_final}.

On the binary detection task, a notable observation is that all models achieve a perfect Precision of 100.0\%. This suggests the non-vulnerable samples in the test set are sufficiently distinct, leading to a near-zero false positive rate. Therefore, the key differentiating metric is Recall. Here, \model\ demonstrates a decisive advantage, achieving a Recall of 97.7\%, which significantly surpasses all baselines by over 34 percentage points. This superiority results in a state-of-the-art F1 Score of 98.8\%.

In the more challenging multi-label CWE detection task, our model continues to show robust performance. It achieves the highest Macro-F1 69.9\% and Micro-F1 70.5\% scores, indicating the most balanced and effective performance overall. This success is primarily driven by its outstanding Macro-Recall and Micro-Recall, once again highlighting its strength in comprehensively identifying all relevant vulnerabilities. While some baselines like \pattern{CodeT5} attain slightly higher Precision, their significantly lower Recall cripples their overall effectiveness.

These strong results validate the capability of our fine-tuned LLM approach for nuanced vulnerability detection. The example in Figure~\ref{fig:example_diversevul} further illustrates our model's capacity to generate accurate and context-aware explanations.

\begin{table*}
\centering
\caption{\revised{Performance Comparison on DiverseVul (RQ1.3). All metric values are percentages (\%). Best results in each row are in \textbf{bold}. * means indicates statistically significant improvement over the baseline method through paired t-test ($p < 0.05$).}}
\renewcommand{\arraystretch}{1.1}
\setlength{\tabcolsep}{6pt}
\begin{tabular}{l l c c c c c}
\toprule
\textbf{Task} & \textbf{Metric} & \textbf{Ours} & \textbf{CodeBERT} & \textbf{CodeT5} & \revised{\textbf{LineVul}} & \revised{\textbf{CausalVul}} \\
\midrule
\multirow{6}{*}{\parbox{2.5cm}{\centering\textbf{Multi-Label CWE Detection}}} & Precision (Micro) & 71.7 & 65.3 & \textbf{72.8} & 65.0 & 67.3 \\
& Recall (Micro) & \textbf{69.3} & 38.2 & 56.8 & 38.1 & 37.4 \\
& F1 Score (Micro) & \textbf{70.5} & 48.0 & 63.9 & 48.0 & 48.3 \\
\cmidrule{2-7}
& Precision (Macro) & 72.5 & 69.0 & \textbf{73.5} & 68.9 & 69.0 \\
& Recall (Macro) & \textbf{67.8} & 32.0 & 52.7 & 32.3 & 32.2 \\
& F1 Score (Macro) & $\textbf{69.9}^*$ & 39.0 & 60.6 & 40.2 & 39.1 \\
\midrule
\multirow{3}{*}{\parbox{2.5cm}{\centering\textbf{Binary Vulnerability Detection}}}
& \revised{Precision} & \textbf{100.0} & \textbf{100.0} & \textbf{100.0} & \textbf{100.0} & \textbf{100.0} \\
& \revised{Recall} & \textbf{97.7} & 61.1 & 62.9 & 63.2 &  60.7 \\
& \revised{F1 Score} & $\textbf{98.8}^*$ & 75.9 & 77.2 & 77.4 & 75.5 \\
\bottomrule
\end{tabular}
\label{table:rq1.3_final}
\end{table*}

\begin{table}[h!]
  \centering
  \begin{tabular}{p{\linewidth}}
  \begin{lstlisting}[language=C++]
  Network::FilterStatus Context::onNetworkNewConnection() {
    onCreate(root_context_id_);
    if (!wasm_->onNewConnection_) {
      return Network::FilterStatus::Continue;
    }
    if (wasm_->onNewConnection_(this, id_).u64_ == 0) {
      return Network::FilterStatus::Continue;
    }
    return Network::FilterStatus::StopIteration;
  }
  \end{lstlisting}
  \begin{mdframed}[style=mystyle]
  \textcolor{red}{[label]} This function is vulnerable. \\
  \textcolor{red}{[cwe]} This function is related to ['CWE-476']. CWE-476 NULL Pointer Dereference: A NULL pointer dereference occurs when the application dereferences a pointer that it expects to be valid, but is NULL, typically causing a crash or exit. \\
  \textcolor{red}{[location]} if (!wasm\_-\textgreater{}onNewConnection\_) \\
  \textcolor{red}{[explanation]} The function is vulnerable to a NULL pointer dereference. It checks if 'wasm\_-\textgreater{}onNewConnection\_' is NULL but does not handle the case when it is NULL. If 'wasm\_-\textgreater{}onNewConnection\_' is NULL, the function continues its execution, which could lead to a NULL pointer dereference and cause a crash or exit. The vulnerability is related to CWE-476: NULL Pointer Dereference.
  \end{mdframed}
  \end{tabular}
  \captionsetup{type=figure}
  \caption{Example of Generated Explanation on DiverseVul.}
  \label{fig:example_diversevul}
\end{table}

\subsubsection{\textbf{RQ1.4: Performance on the PrimeVul Dataset}}

\phead{Task Setup.}
To further probe the limits of current models, we conduct a more stringent evaluation using the \pattern{PrimeVul} dataset. We This task is formulated as multi-class classification, where each vulnerable function belongs to exactly one CWE type. \pattern{PrimeVul} is recognized as a more challenging benchmark due to its wider variety of vulnerability types and more realistic, imbalanced data distribution, which rigorously tests a model's ability to distinguish between fine-grained vulnerability patterns.

\phead{Experimental Results.}
The performance on \pattern{PrimeVul}, detailed in Table~\ref{table:rq1.4_final}, presents a complex and challenging picture that highlights the critical issue of class imbalance in vulnerability detection.

In the multi-class CWE detection task, the high Weighted-F1 scores (e.g., 94.9\% for \model) initially suggest strong performance. However, these scores are misleading as they are dominated by the models' accuracy on the major non-vulnerable code. The performance of all models plummets on the Macro-averaged metrics, with the top Macro-F1 score being a mere 21.7\% from \pattern{CausalVul}, and our model achieving a comparable 20.7\%. This dramatic gap between Weighted and Macro scores proves that all models, while able to learn common vulnerability patterns, largely fail to generalize to the long tail of less frequent CWEs—a critical failure for real-world applications.

This difficulty is further confirmed by the binary detection task. On this more challenging dataset, all models experienced a drastic performance drop compared to their results on \pattern{DiverseVul}. Our model, \model, while achieving the highest F1-Score at 39.8\%, still operates at a level far below what is required for reliable, practical use.

The findings from \pattern{PrimeVul} strongly suggest that current fine-tuning methodologies for LLMs, despite their promise, are not yet robust enough for complex, imbalanced, real-world vulnerability analysis. This underscores a critical research gap and motivates the need for more advanced techniques to enhance model generalization and robustness against low-frequency data.

\begin{table*}
\centering
\caption{\revised{Performance Comparison on PrimeVul (RQ1.4). All metric values are percentages (\%). Best results in each row are in \textbf{bold}. * means indicates statistically significant improvement over the baseline method through paired t-test ($p < 0.05$).}}
\renewcommand{\arraystretch}{1.1}
\setlength{\tabcolsep}{6pt}
\begin{tabular}{l l c c c c c}
\toprule
\textbf{Task} & \textbf{Metric} & \textbf{Ours} & \textbf{CodeBERT} & \textbf{CodeT5} & \revised{\textbf{LineVul}} & \revised{\textbf{CausalVul}} \\
\midrule
\multirow{6}{*}{\parbox{2.5cm}{\centering\textbf{Multi-Class CWE Detection}}} & Precision (Weighted) & \textbf{96.2} & 96.0 & 95.9 & 95.2 & 96.1\\
& Recall (Weighted) & 93.1 & 90.2 & 89.2 & \textbf{93.4} & 90.4\\
& F1 Score (Weighted) & \textbf{94.9} & 93.1 & 91.9 & 94.1 & 93.3\\
\cmidrule{2-7}
& Precision (Macro) & 19.2 & 13.7 & 18.7 & 14.4 & \textbf{25.0}\\
& Recall (Macro) & \textbf{25.7} & 20.1 & 20.4 & 19.3 & 19.2\\
& F1 Score (Macro) & 20.7 & 14.1 & 15.3 & 14.0 & \textbf{21.7}\\
\midrule
\multirow{3}{*}{\parbox{2.5cm}{\centering\textbf{Binary Vulnerability Detection}}}
& \revised{Precision} & \textbf{29.5} & 19.6 & 18.2 & 25.8 & 20.2 \\
& \revised{Recall} & \textbf{61.1} & 60.1 & \textbf{61.1} & 47.8 & 59.4 \\
& \revised{F1 Score} & $\textbf{39.8}^*$ & 29.6 & 28.0 & 33.5 & 30.1 \\
\bottomrule
\end{tabular}
\label{table:rq1.4_final}
\end{table*}

\greybox{Summary of RQ1}{Our evaluation offers a dual perspective on LLMs for vulnerability detection. It confirms their feasibility and effectiveness on established benchmarks like \pattern{SeVC} and \pattern{DiverseVul}. However, it also reveals their current limitations, as all models struggled with the class imbalance and complexity of the more challenging \pattern{PrimeVul} dataset, highlighting a critical area for future research.}

\subsection{\textbf{RQ2: How proficient are LLMs in explaining the detected vulnerabilities?}}

In this RQ, we explore the proficiency of the \model in explaining the detected vulnerabilities. While numerous evaluation metrics exist for assessing text generation by LLMs~\cite{papineni2002bleu,zhang2019bertscore}, there is currently no comprehensive framework tailored to evaluating vulnerability explanations. The explanatory content generated by the model should assist software developers in identifying and mitigating potential vulnerabilities by ensuring the accuracy of the analysis, maintaining readability and conciseness, and providing actionable suggestions for remediation.

The evaluation of explanation quality remains an open problem, as there are no definitive standards for what constitutes a ``correct'' or ``complete'' explanation. This inherent subjectivity poses significant challenges for effective evaluation. Therefore, the evaluation of explanations must be contextualized within the specific goals of the vulnerability explanation task. By summarizing and generalizing the “helpfulness” of the generated text, we assess explanation quality from three critical perspectives: \textbf{Accuracy}, \textbf{Clarity}, and \textbf{Actionability}.

\subsubsection{\textbf{Experimental Setup.}} \par \noindent
In this study, we evaluate the explanations generated by the fine-tuned \model for vulnerability code from the \pattern{DiverseVul} and \pattern{PrimeVul} test sets, alongside ground truth annotations generated using Deepseek-V3.
Given the large scale of the dataset and the high cost of evaluation, we employ a random sampling approach based on a 95\% confidence level and a 5\% confidence interval~\cite{sample}. This yields 224 and 197 samples from \pattern{DiverseVul} and \pattern{PrimeVul}, respectively, ensuring statistical validity and representativeness in our evaluation.

Our evaluation includes two parts:
{\em \textbf{(1) Manual review}}: Two authors, with extensive research experience in software vulnerabilities and development, independently review the sampled data, scoring them based on predefined criteria. Disagreements are resolved through discussion until a consensus is reached. The Cohen's Kappa values~\cite{kappa} for inter-reviewer agreement are 0.78 and 0.81 for two datasets, indicating substantial consistency and reliability.
{\em \textbf{(2) Automated review using LLM:}} Deepseek-V3 is used for automated evaluation. A detailed prompt is provided, describing the vulnerability explanation evaluation task and the specific criteria. The LLM is instructed to output scores for the three evaluation metrics, with 1 indicating satisfaction and 0 indicating non-satisfaction.

This dual evaluation approach provides a robust assessment of the quality of the explanations generated by the fine-tuned model and demonstrates the feasibility of using LLMs for scalable evaluation tasks in the future.

\phead{Evaluation Metrics.} 
The three evaluation metrics used in this study are defined as follows:
\begin{itemize}[leftmargin=1.5em]
    \item \textbf{Accuracy}: The explanation should correctly identify and describe the vulnerability, ensuring the details are factually correct and relevant to the detected vulnerability.
    \item \textbf{Clarity}: The explanation should be clear, concise, and structured in a way that facilitates easy comprehension by software developers.
    \item \textbf{Actionability}: The explanation include actionable suggestions for remediation, offering practical guidance for addressing the identified vulnerability.
\end{itemize}

\subsubsection{\textbf{Experimental Results.}} \par \noindent
We present the results of our dual-rater (Manual and LLM-Automated) evaluation in Table~\ref{table:RQ2}. The column suffixes denote the data source: {\sf -Gen} refers to explanations generated by our model (\model), while {\sf -Ann} refers to the higher-quality explanations produced by our annotation method. {\sf All-Pos.} is a strict metric representing cases that satisfy all three criteria (Accuracy, Clarity, and Actionability). Our analysis yields three key findings:

\phead{(1) Generated Explanations Show Value, but Annotated Explanations Are Superior.}
Our model-generated explanations ({\sf -Gen}) demonstrate practical value, achieving, for instance, a manual All-Positive score of 51.3\% on \pattern{DiverseVul}. However, the results consistently show that the explanations produced via our annotation method ({\sf -Ann}) are of significantly higher quality across all metrics and datasets. For example, on \pattern{PrimeVul}, the manually-assessed All-Positive score for {\sf PRI-Ann} (90.3\%) is more than double that of {\sf PRI-Gen} (41.6\%). This validates the effectiveness of our annotation pipeline in creating a high-quality dataset for fine-tuning and evaluation. The lower scores on \pattern{PrimeVul} compared to \pattern{DiverseVul} also echo our findings from RQ1, confirming it as a more challenging benchmark for both detection and explanation.

\phead{(2) Agreement Varies Significantly Between Objective and Subjective Metrics.}
The Cohen's Kappa coefficients, which measure the agreement between manual and LLM raters, reveal a critical insight. For the relatively objective \textbf{Accuracy} metric, which assesses factual correctness (e.g., correct location, valid CWE), the agreement is consistently in the moderate to substantial range (Kappa from 0.52 to 0.65). This indicates that LLMs can reliably replicate human judgment when the evaluation criteria are concrete and verifiable.

In stark contrast, the agreement for the more subjective metrics of \textbf{Clarity} and \textbf{Actionability} is significantly lower and more varied. On \pattern{DiverseVul}, the Kappa values are near zero (0.01 to 0.12), signaling almost no agreement beyond chance. While the agreement is better on \pattern{PrimeVul}, it remains modest. This discrepancy strongly suggests that while LLMs are proficient at automating factual checks, they struggle to consistently align with human assessment of subjective qualities like writing style, comprehensibility, and the perceived usefulness of repair suggestions.

\phead{(3) LLMs Are Viable for Automating Concrete, Fact-Based Evaluation.}
Based on the Kappa analysis, we conclude that LLMs are most suitable for automating the evaluation of specific, clearly-defined aspects of an explanation's quality. 
Our comparative analysis of inter-rater agreement reveals a clear and practical boundary for using LLMs in evaluation. Compared to human-human agreement, the human-LLM agreement is relatively lower. Crucially, this significant gap is not uniform. The discrepancy is primarily driven by the LLM's struggle with nuanced, subjective metrics like Clarity and Actionability.
The strong agreement on the `Accuracy` metric validates the potential of LLM-based automation for reliably checking the factual correctness of generated explanations at scale. This allows human experts to focus their limited time on the more nuanced, subjective aspects that still require human oversight. Our dual-validation strategy, therefore, offers a practical framework: leveraging LLMs for scalable, objective assessment while using targeted human review for subjective quality control, creating an efficient and reliable evaluation pipeline.

While the proposed approach demonstrates strong performance, it has inherent limitations. Fully correct explanations cannot be guaranteed due to the complexity of vulnerability analysis, and manual evaluations are subject to reviewer bias. Despite these challenges, the dual-validation strategy combining manual and automated reviews offers a practical and effective solution for evaluating vulnerability explanations, balancing accuracy with scalability and cost-effectiveness.

\begin{table}
    \centering
    \caption{Proportion (\%) of Vulnerability Explanation Review Results and Inter-Rater Agreement (RQ2).}
    \resizebox{\columnwidth}{!} {
    \begin{tabular}{l|c|c|c|c}
        \hline
        \textbf{Metric} & \textbf{DIV-Gen.} & \textbf{DIV-Ann.} & \textbf{PRI-Gen.} & \textbf{PRI-Ann.} \\ \hline
        \multicolumn{5}{c}{\textbf{Manual}} \\ \hline
        \textbf{Accuracy} & 57.4 & 84.3 & 43.8 & 93.0 \\ \hline
        \textbf{Clarity} & 96.7 & 99.0 & 71.4 & 97.8 \\ \hline
        \textbf{\revised{Action.}} & 82.7 & 83.2 & 42.7 & 90.3 \\ \hline
        \multicolumn{1}{>{\columncolor[gray]{0.90}}l|}{\textbf{All-Pos.}} & \multicolumn{1}{>{\columncolor[gray]{0.90}}c|}{51.3} & \multicolumn{1}{>{\columncolor[gray]{0.90}}c|}{71.1} & \multicolumn{1}{>{\columncolor[gray]{0.90}}c|}{41.6} & \multicolumn{1}{>{\columncolor[gray]{0.90}}c}{90.3} \\ \hline
        \multicolumn{5}{c}{\textbf{LLM-Automation}} \\ \hline
        \textbf{Accuracy} & 62.9 & 89.3 & 46.5  & 94.6 \\ \hline
        \textbf{Clarity} & 90.9 & 97.5 & 73.5 & 99.5 \\ \hline
        \textbf{\revised{Action.}} & 61.4 & 85.3 & 47.0 & 93.5 \\ \hline
        \multicolumn{1}{>{\columncolor[gray]{0.90}}l|}{\textbf{All-Pos.}} & \multicolumn{1}{>{\columncolor[gray]{0.90}}c|}{59.9} & \multicolumn{1}{>{\columncolor[gray]{0.90}}c|}{85.3} & \multicolumn{1}{>{\columncolor[gray]{0.90}}c|}{46.5} & \multicolumn{1}{>{\columncolor[gray]{0.90}}c}{93.0} \\ \hline

        \multicolumn{5}{c}{\revised{\textbf{Agreement (Cohen's Kappa)}}} \\ \hline
        \multicolumn{5}{l}{\revise{\textit{Human vs. Human Agreement (per-metric)}}} \\ \hline 
        \revise{\textbf{Accuracy}} & \revise{0.92} & \revise{0.90} & \revise{0.77} & \revise{0.75} \\
        \revise{\textbf{Clarity}} & \revise{0.76} & \revise{0.66} & \revise{0.82} & \revise{0.85} \\
        \revise{\textbf{Actionability}} & \revise{0.60} & \revise{0.60} & \revise{0.64} & \revise{0.65} \\ \hline

        \multicolumn{5}{l}{\revised{\textit{Human vs. LLM Agreement (per-metric)}}} \\ \hline 
        \revised{\textbf{Accuracy}} & \revised{0.65} & \revised{0.52} & \revised{0.60} & \revised{0.58} \\
        \revised{\textbf{Clarity}} & \revised{0.04} & \revised{0.01} & \revised{0.57} & \revised{0.39} \\
        \revised{\textbf{Actionability}} & \revised{0.02} & \revised{0.12} & \revised{0.54} & \revised{0.57} \\ \hline

        \multicolumn{5}{c}{\revised{\textbf{Overall Agreement Comparison}}} \\ \hline 
        \revise{\textbf{Human vs. Human}} & \multicolumn{2}{c|}{\revise{\textbf{0.78}}} & \multicolumn{2}{c}{\revise{\textbf{0.81}}} \\ \hline
        \revise{\textbf{Human vs. LLM}} & \multicolumn{2}{c|}{\revise{\textbf{0.38}}} & \multicolumn{2}{c}{\revise{\textbf{0.68}}} \\ \hline

    \end{tabular}
    }
    \label{table:RQ2}
\end{table}

\subsubsection{\revised{\textbf{Qualitative Analysis of Vulnerability Explanations}}} \par \noindent

Beyond the numerical scores, a qualitative analysis of the generated explanations is crucial to understand why the models perform as they do. By examining both exemplary and unsatisfactory cases, we can uncover the models' underlying capabilities and current limitations.

\phead{An Exemplary Case.}
To highlight the importance of evaluating vulnerability explanations, we analyze an example from \pattern{SeVC}. The sample pertains to a pointer-type vulnerability caused by incorrect memory deallocation in C code. As shown in Figure~\ref{table:case}, the explanation correctly identifies the issue: the delete operator is used to free stack-allocated memory created by the ALLOCA function, which should instead use the free function. The explanation provides actionable suggestions, such as ensuring pointer safety by using the appropriate deallocation methods. It is clear, logically structured, and actionable, satisfying all three evaluation criteria.

\begin{table}[h!]
  \centering
  \vspace{-0.2cm}
  \begin{tabular}{p{\linewidth}}
    \begin{lstlisting}[language=C++]
int i;
for(i = 0; i < 1; i++)
twoIntsStruct * dataBuffer = (twoIntsStruct *) ALLOCA(100 * sizeof(twoIntsStruct));
size_t i;
for (i = 0; i < 100; i++)
  dataBuffer[i].intOne = 1;
  dataBuffer[i].intTwo = 1;
data = dataBuffer;
printStructLine(&data[0]);

void printStructLine(const twoIntsStruct * structTwoIntsStruct) {
  printf(``%d -- %d\n'', structTwoIntsStruct->intOne, structTwoIntsStruct->intTwo);
}
delete [] data;
    \end{lstlisting} \\
    \vspace{-1cm}
    \begin{mdframed}[style=mystyle]
    \textcolor{red}{[type]} pointer \\
    \textcolor{red}{[location]} The line ``delete [] data'' has a pointer-related issue. \\
    \textcolor{red}{[explanation]} \\
    \textcolor{blue}{(Analysis:)}
    The issue in this code is related to the ``data'' pointer, which is being improperly freed using the ``delete'' operator. The ``data'' pointer is assigned the address of a memory block allocated on the stack using the ``ALLOCA'' function, which allocates memory on the stack instead of the heap. The ``delete'' operator is used to free memory allocated on the heap using the ``new'' operator, and using it to free memory allocated on the stack may cause undefined behavior or crashes. \\
    \textcolor{blue}{(Suggestion:)}
    To fix this issue, the program should use the ``free'' function to free memory allocated on the stack, or use the ``new'' operator to allocate memory on the heap instead of the ``ALLOCA'' function. Additionally, the program should ensure that the ``data'' pointer points to a valid memory location before it is used or accessed, and should add proper error handling and validation to ensure that the ``data'' pointer behaves correctly and safely in all cases.
    \end{mdframed} \\
  \end{tabular}
      \captionsetup{type=figure}
  \caption{Example of Generated Explanation on SeVC.}
  \label{table:case}
\end{table}

\phead{Insights from Unsatisfactory Explanations.}
An analysis of the less satisfactory explanations reveals recurring patterns that not only explain the numerical results in Table~\ref{table:RQ2} but also offer crucial insights for future research.

\noindent \textbf{1. Factual Errors vs. Subjective Deficiencies: Explaining the Kappa Discrepancy.}
The shortcomings in generated explanations can be broadly categorized into two types, which directly explain the variance in inter-rater agreement (Kappa scores).

Factual Inaccuracies: The first category includes objective, verifiable errors. These are the primary reason for low Accuracy scores and include: (a) Incorrect Localization, where the model identifies the wrong line number; (b) Root Cause Misidentification, where it describes a symptom (e.g., a crash) instead of the underlying flaw (e.g., a use-after-free); and (c) Vulnerability Misclassification. Because these errors are factual, both human and LLM raters can assess them with high consistency, leading to the moderate-to-substantial Kappa agreement (0.52-0.65) observed for the Accuracy metric.

Subjective Deficiencies: The second category involves subjective qualities of the explanation. These include: (a) Generic or Missing Repair Guidance, where suggestions are too vague (e.g., ``validate all inputs''); and (b) Poor Clarity, due to excessive verbosity or templated, repetitive phrasing. Assessing these qualities is inherently subjective—what one developer finds clear, another may find verbose. This subjectivity is the primary reason for the low-to-minimal Kappa agreement seen for Clarity and Actionability, especially on the \pattern{DiverseVul} dataset.

\noindent \textbf{2. Shallow Pattern Matching vs. Deep Semantic Reasoning: The Core Limitation.}
A deeper insight is that many of these failures stem from a single core limitation: the model often relies on shallow pattern matching rather than deep, causal reasoning about the code's execution flow.

This hypothesis explains several observed phenomena. Incorrect Localization often results from matching superficial tokens (e.g., a function name) without understanding the control flow that activates the vulnerability. The model's tendency to produce templated or generic explanations suggests it is retrieving and adapting a learned pattern rather than reasoning from first principles. Most importantly, this explains why performance was notably worse on the more complex \pattern{PrimeVul} dataset (Table~\ref{table:RQ2}). The vulnerabilities in \pattern{PrimeVul} likely require a more global understanding of inter-procedural dependencies, which is a known weakness of models that rely on local, pattern-based reasoning. The model excels at explaining common vulnerabilities it has seen frequently during training but struggles when faced with novelty or complexity.

\phead{Implications.}
The qualitative analysis reveals the dual nature of current LLMs for vulnerability explanation. They are powerful pattern recognizers, capable of producing excellent explanations for common vulnerabilities in relatively simple code. However, their reliability diminishes when faced with complex or novel scenarios that require deep semantic understanding. This underscores that while LLMs are invaluable assistive tools, they are not yet infallible replacements for human expertise. The path forward for research is clear: we must develop methods that move beyond pattern matching to instill a more robust, causal reasoning capability in these models.

\greybox{Summary of RQ2}{Our analysis reveals a critical insight for automated explanation evaluation: LLMs show strong agreement with human experts on factual Accuracy but diverge significantly on subjective metrics like Clarity and Actionability. This suggests that the most promising path forward is a hybrid evaluation strategy, where LLMs automate scalable, objective checks, while human oversight remains essential for nuanced, subjective quality assessment.}

\subsection{\textbf{RQ3: How do explanations affect the results of vulnerability detection?}}

Building on the detection capabilities evaluated in RQ1, this section investigates a critical question: Does the act of generating explanations help or hinder the primary task of vulnerability detection?
To answer this, we experimentally examine the impact of the explanation task by comparing our standard model, which is fine-tuned to both detect and explain, against a variant fine-tuned solely for detection.

\phead{Experimental Setup.}
We conduct this ablation study on the \pattern{SeVC}, \pattern{DiverseVul} and \pattern{PrimeVul} datasets. To isolate the effect of the explanation task, we create a detection-focused counterpart for our model. This is achieved by removing all explanatory text from the training data and adjusting the instruction prompts so that the model's only goal is to output the vulnerability label(s). We refer to our standard model as the Explainer and its detection-only counterpart as the Detector. All other hyperparameters and training procedures are kept identical.

\begin{figure}[t]
 \centering
 \includegraphics[width=0.92\linewidth]{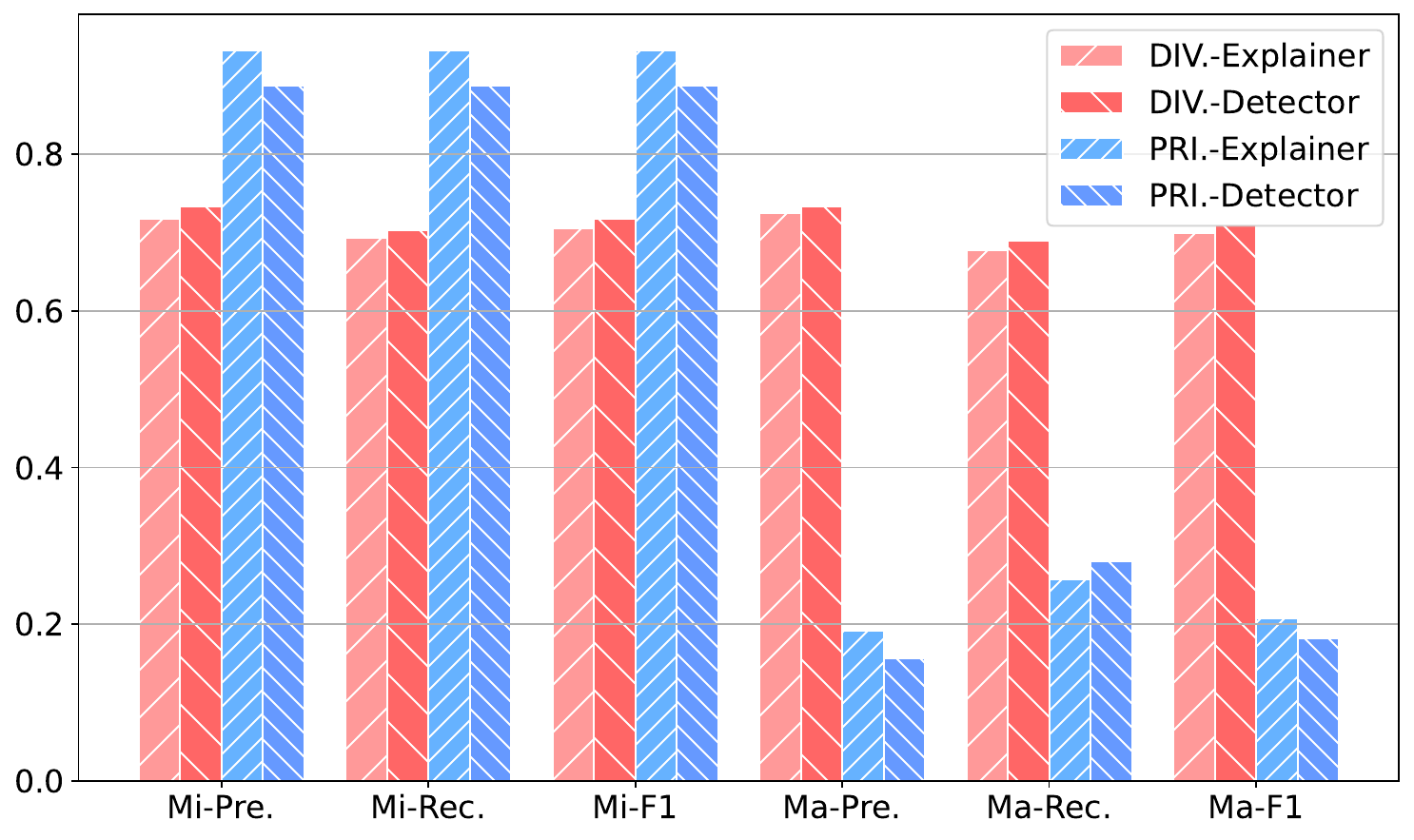}
 \caption{Performance comparison of the Explainer (trained to explain and detect) and the Detector (trained only to detect) on the DiverseVul and PrimeVul datasets.}
 \label{fig:rq3}
\end{figure}

\phead{Experimental Results.}
We first show the comparison of Detector and Explainer on easier SeVC for multi-type vulnerability detection in Table~\ref{tab:rq3_sevc}. Our experimental results reveal that detection-only models outperform interpretable models on the SeVC synthetic dataset. This occurs because synthetic vulnerability patterns in SeVC are highly discernible, and interpretation task training directs the model's focus toward explanation generation, slightly compromising detection performance. The marginal performance discrepancy indicates strong compatibility between these tasks, demonstrating that interpretation tasks can jointly participate in large model fine-tuning alongside detection objectives."

\begin{table}
\centering
\caption{SeVC Dataset Experiment Results Comparison}
\label{tab:rq3_sevc}
\resizebox{\columnwidth}{!} {
\begin{tabular}{lccccccc}
\toprule
         Model &  Mi-Pre. &  Mi-Rec. &  Mi-F1 &  Ma-Pre. &  Ma-Rec. &  Ma-F1 \\
\midrule
SeVC-Explainer &   79.9 &   80.5 & 80.1 &   70.0 &   72.4 & 70.8 \\
SeVC-Detector &   81.7 &   81.4 & 81.2 &   70.6 &   75.4 & 72.0 \\
\bottomrule
\end{tabular}
}
\end{table}

The comparative results of real-world vulnerability detection, shown in Figure~\ref{fig:rq3}, reveal a nuanced and dataset-dependent relationship between explanation and detection performance. The impact of learning to explain is not monolithic; instead, it is dictated by the complexity of the underlying task.

On the simpler benchmark (\pattern{DiverseVul}), the detection-only model (`Detector`) shows a marginal performance advantage. For instance, the `Detector` achieves a slightly higher Macro-F1 score than the `Explainer` (71.0\% vs. 69.9\%). This suggests that on less complex tasks, forcing the model to also generate explanations may introduce a minor multi-task learning overhead, slightly diluting its focus on the primary detection objective.
{On the more challenging benchmark (\pattern{PrimeVul}), this trend dramatically reverses. Here, the `Explainer` model significantly outperforms the `Detector` across nearly all metrics. The most notable advantage is seen in the Micro-F1 score (93.2\% vs. 88.7\%), but the trend also holds for the more challenging Macro-F1 score (20.7\% vs. 18.2\%). This reversal is a critical finding. It strongly suggests that for complex, real-world vulnerabilities with subtle patterns, the task of generating explanations is not a hindrance. Instead, it acts as a form of implicit regularization or deep supervision.
By learning to articulate why a piece of code is vulnerable, the model is compelled to develop a more robust and granular understanding of the vulnerability's root cause. This deeper semantic comprehension, forged through the explanation task, directly enhances its ability to detect those same complex vulnerabilities, providing a performance benefit that far outweighs any multi-task overhead.

These findings demonstrate that the vulnerability explanation task can coexist with the detection task without compromising the model's detection accuracy. The Explainer retains strong detection performance while gaining additional explanatory capabilities. Furthermore, automatically annotated explanatory data introduces domain-specific knowledge, enhancing the model's ability to understand and identify diverse vulnerability patterns.
In summary, enhancing a model's explanatory capabilities does not come at the cost of detection performance. In some cases, well-annotated explanatory data can even improve the model’s overall understanding and effectiveness across both tasks.

\greybox{Summary of RQ3}{The relationship between explanation and detection is nuanced and depends on task complexity. For simpler vulnerabilities, a detection-only focus is marginally better. For complex, realistic vulnerabilities, however, fine-tuning for explanation significantly boosts detection performance by fostering a deeper, more robust understanding of root causes.}

\subsection{\textbf{RQ4: How does the key code extraction impact detection and explanation} performance?}

In this RQ, we investigate whether explicitly guiding an LLM to focus on critical code segments can enhance its performance. To explore this, we creat datasets for \pattern{SeVC}, \pattern{DiverseVul} and \pattern{PrimeVul} where likely vulnerable code snippets (``key code'') are pre-annotated. We then fine-tune and evaluate our model using this focused information.

\phead{Experimental Setup.}
As outlined in Section~\ref{sec:methodology}, we employ LLMs to annotate key code segments related to vulnerabilities in our datasets. To prevent data leakage, vulnerability type information is concealed during this annotation process. During the fine-tuning and inference stages for the ``With Key Code'' model, our prompts are modified to instruct the model to specifically leverage these extracted snippets for its analysis. We then compare this model's detection performance against the baseline model trained ``Without Key Code''. We also briefly assess the impact on explanation quality using the same methodology as in RQ2.

\begin{table}
\centering
\small
\caption{F1 of Key Code Extract. on SeVC: (RQ4).}
\resizebox{\columnwidth}{!} {
\tabcolsep=8pt
\renewcommand{\arraystretch}{0.95}
\begin{tabular}{c|cc|cc}
\hline
\multirow{2}{*}{Type} & \multicolumn{2}{c|}{Single-Type(RQ1.1)} & \multicolumn{2}{c}{Multi-Type(RQ1.2)} \\ \cline{2-5} 
 & W/O Key. & With Key. & W/O Key. & With Key. \\ \hline
API & 93.0\% & \cellcolor[HTML]{EFEFEF}\textbf{98.7\%} & 79.8\% & \cellcolor[HTML]{EFEFEF}\textbf{87.8\%} \\ 
Arith. & 91.9\% & \cellcolor[HTML]{EFEFEF}\textbf{97.9\%} & 71.9\% & \cellcolor[HTML]{EFEFEF}\textbf{96.0\%} \\
Pointer & 92.4\% & \cellcolor[HTML]{EFEFEF}\textbf{99.1\%} & 58.3\% & \cellcolor[HTML]{EFEFEF}\textbf{73.5\%} \\ 
Array & 92.4\% & \cellcolor[HTML]{EFEFEF}\textbf{98.2\%} & 47.4\% & \cellcolor[HTML]{EFEFEF}\textbf{77.9\%} \\ 
Average & 92.4\% & \cellcolor[HTML]{EFEFEF}\textbf{98.5\%} & 64.4\% & \cellcolor[HTML]{EFEFEF}\textbf{83.8\%} \\ \hline
\end{tabular}
}
\label{table:rq4.1}
\end{table}

\begin{table}[t]
\centering
\caption{Impact of Key Code on Detection Performance on the \pattern{PrimeVul} Dataset (RQ4).}
\label{table:primevul_keycode_ablation}
\renewcommand{\arraystretch}{1.1}
\setlength{\tabcolsep}{10pt}
\begin{tabular}{l c c}
\toprule
\textbf{Metric} & \textbf{Without Key Code} & \textbf{With Key Code} \\
\midrule
\multicolumn{3}{l}{\textbf{Multi-Class CWE Detection}} \\
\cmidrule(l){1-3}
Precision (Macro)    & 19.2 & \textbf{19.3} \\
Recall (Macro)       & 25.7 & \textbf{27.6} \\
F1 Score (Macro)     & 20.7 & \textbf{21.5} \\
\midrule
\multicolumn{3}{l}{\textbf{Binary Vulnerability Detection}} \\
\cmidrule(l){1-3}
Precision            & 29.5 & \textbf{31.7} \\
Recall               & 61.1 & \textbf{62.6} \\
F1 Score             & 39.8 & \textbf{42.1} \\
\bottomrule
\end{tabular}
\end{table}

\begin{figure}[t]
 \centering
 \includegraphics[width=0.92\linewidth]{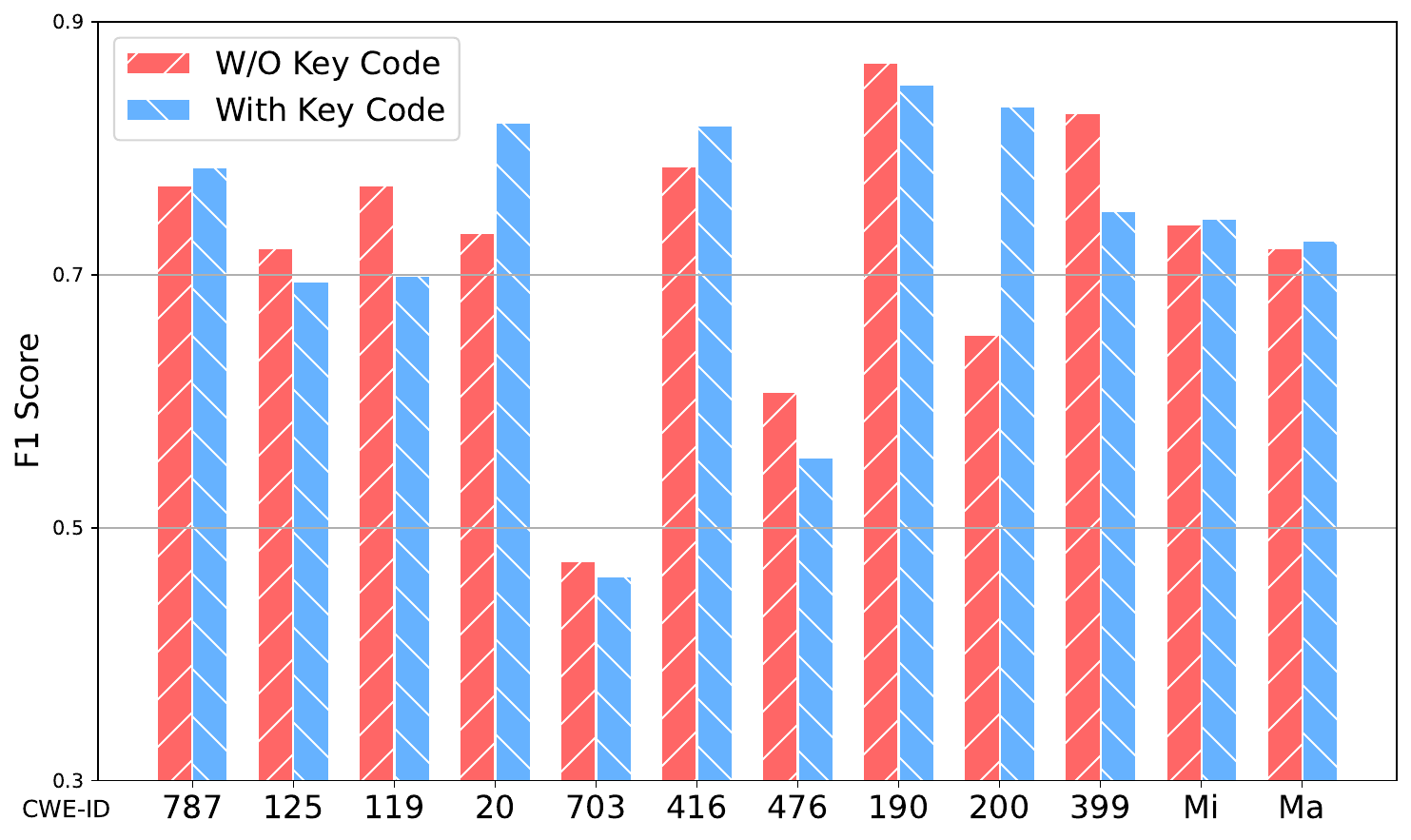}
 \caption{Impact of Key Code on Detection Performance on the \pattern{DiverseVul} Dataset (RQ4).}
 \label{fig:rq4.2}
 \end{figure}

\phead{Experimental Results.}
Our findings indicate that the effectiveness of key code extraction is highly dependent on the complexity of the dataset.

On \pattern{SeVC}—the dataset featuring synthesized data and relatively simple vulnerability classification—key code extraction yields notably superior performance as demonstrated in Table~\ref{table:rq4.1}.
On the more challenging \pattern{PrimeVul} dataset, key code provides a clear advantage. As shown in Table~\ref{table:primevul_keycode_ablation}, the ``With Key Code'' model consistently outperforms the baseline. The F1 score for binary detection improves from 39.8\% to 42.1\%, and the Macro-F1 for multi-class detection increases from 20.7\% to 21.5\%. This positive result is significant because \pattern{PrimeVul} is the benchmark where models struggled most (RQ1.4). The improvement stems from two benefits: (1) \textbf{Enhanced Focus}, which directs the model's attention to the most relevant code, and (2) \textbf{Noise Reduction}, which improves the signal-to-noise ratio by filtering out irrelevant context. This demonstrates that for complex, noisy, real-world code, guiding the model's focus is a highly effective strategy.

On the \pattern{DiverseVul} dataset, the impact is inconclusive. In contrast, the results illustrated in Figure~\ref{fig:rq4.2} show a mixed trend, with neither model establishing a definitive advantage. A plausible explanation is that the vulnerabilities in \pattern{DiverseVul} are more straightforward, allowing the baseline model to perform well without explicit guidance. In such cases, the overhead of the key code extraction process offers no significant additional benefit.

Regarding explanation quality, our manual and LLM evaluation investigation that simply providing key code during fine-tuning did not yield a significant improvement in the quality of generated explanations. This suggests that a more sophisticated integration strategy is needed to translate focused attention into higher-quality explanatory text. Exploring how to better leverage key code to enhance explanation quality remains an important direction for future research.

\greybox{Summary of RQ4}{Key code extraction enhances vulnerability detection performance on challenging real-world datasets by effectively guiding the model's attention mechanisms.}
\section{Discussion}
\label{sec:discussion}
In this section, we discuss the implications of our study.

\phead{Implication 1: Broader Applications for Vulnerability-Related Tasks.}
Our finding that integrating vulnerability explanation with detection preserves model performance suggests expanding LLM fine-tuning beyond detection. Incorporating tasks like impact assessment, severity prioritization, or automated patch generation could enhance contextual understanding and actionability. This holistic approach may yield tools that simultaneously identify and mitigate vulnerabilities.

\phead{Implication 2: Effectiveness of \model Using Different LLMs.}
In this paper, we use \pattern{CodeLlama-13B-Instruct} as the primary model to conduct the experiments.
A key question is how our framework's effectiveness varies with different base LLMs. We investigate this from both smaller and larger directions.

First,  we utilize two additional widely-used smaller LLMs, \pattern{CodeLlama-7B-Instruct} and \pattern{Llama3-8B-Instruct}, on the easier DiverseVul dataset. As shown in Figure~\ref{fig:DS_2}, \model achieves effective and comparable performance with these models, demonstrating that our proposed fine-tuning and explanation generation strategies are not limited to a single model architecture and can be successfully applied to smaller LLMs.
Second, we seek to understand if simply employing a larger, more advanced model could further boost performance, particularly on a more challenging, realistic dataset. To this end, we conduct experiments on the PrimeVul dataset, comparing the performance of \pattern{CodeLlama-13B-Instruct} against a significantly larger and more powerful model, \pattern{DeepSeek-Coder-33B-Instruct}. The results, presented in Figure~\ref{fig:DS_prime}, lead to a crucial insight: the more advanced model did not yield a discernible accuracy gain.
This finding consistently suggests that for fine-tuning tasks on current, small-scale vulnerability datasets, simply increasing the base model's size or using a more recent architecture may not be the silver bullet. We hypothesize that the primary bottleneck for improving performance is now the scale and diversity of the training data, rather than the inherent capacity of the base model. This is a very valuable insight for the field, suggesting that future efforts should focus on data-centric approaches to unlock the full potential of large models for vulnerability detection.

\begin{figure}
 \centering
\includegraphics[width=0.92\linewidth ]{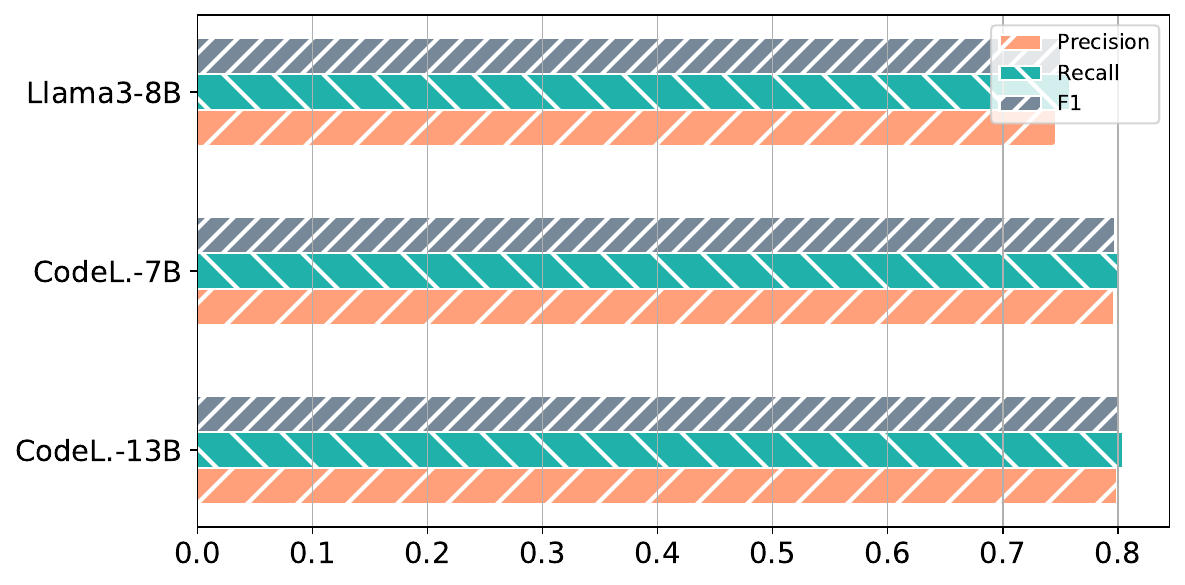}
 \caption{Different base LLMs on DiverseVul.}
 \label{fig:DS_2}
 \end{figure}

 \begin{figure}
 \centering
\includegraphics[width=0.92\linewidth ]{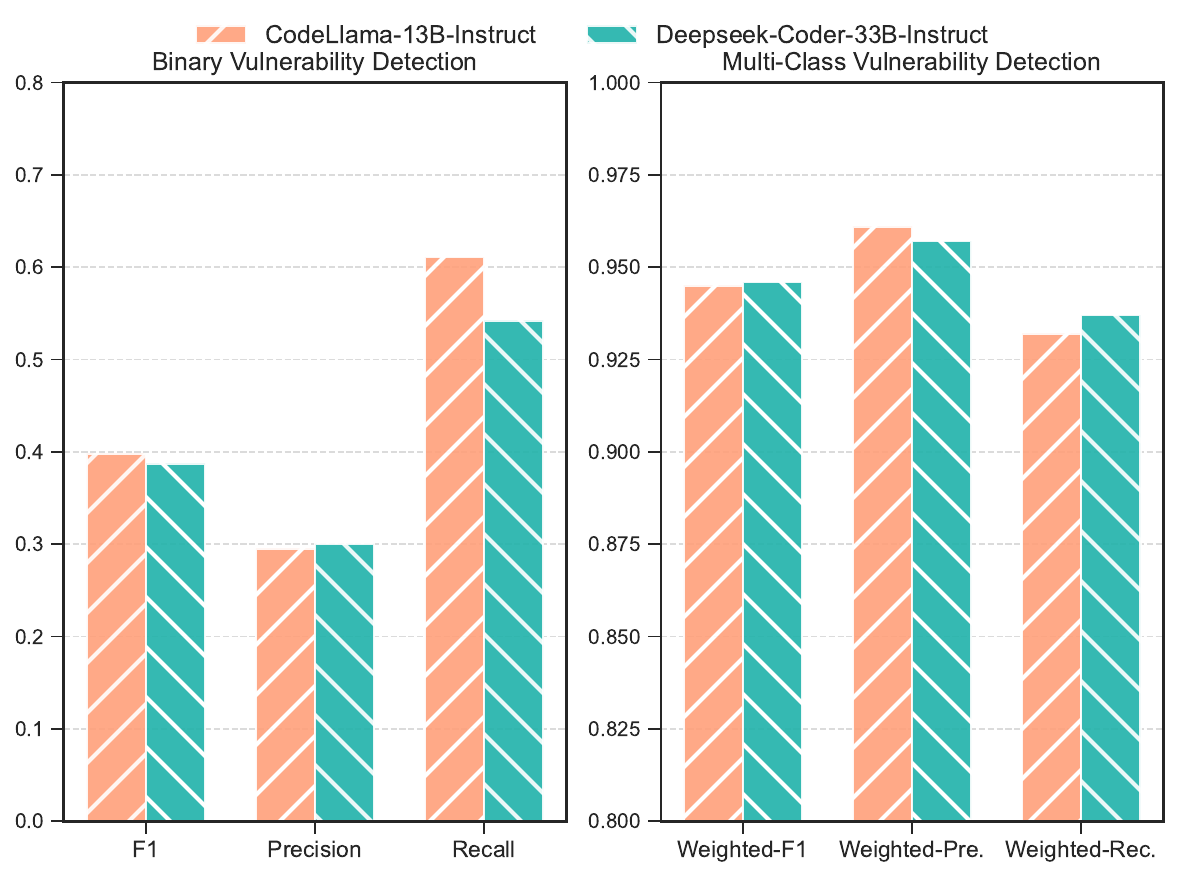}
 \caption{Different base LLMs on PrimeVul.}
 \label{fig:DS_prime}
 \end{figure}

\phead{Implication 3: Dependency on Annotation Quality in Fine-Tuning Frameworks.}
To demonstrate the high sensitivity of fine-tuning effectiveness to annotation quality, we conduct experiments using different annotation LLMs. Specifically, CodeLlama-7B fine-tuned with DeepSeek-V3-generated explanations achieved an average Macro-F1 of 20.5\% on PrimeVul, while the same model fine-tuned with CodeLlama-13B-generated explanations attained only 16.4\%. This significant performance gap substantiates the critical role of annotation quality.
This dependency underscores the necessity of developing robust annotation frameworks that ensure consistency and accuracy. A promising approach involves semi-automated tools where advanced LLMs provide initial annotations subsequently refined by human experts. Such a hybrid methodology could mitigate quality variations while balancing the practical constraints of manual curation.
\section{Threats to Validity}
\label{sec:threats}

\phead{Construct Validity.}
We evaluated explanatory capability using three metrics (Accuracy, Clarity, and Actionability). While informative, these metrics may not be exhaustive, and further refinement is desirable. To mitigate bias, two authors independently annotated explanations, achieving substantial agreement. Furthermore, our use of GPT-3.5 and DeepSeek-V3 for certain automated annotation tasks introduces a replication challenge: its specific version may deprecate or evolve, potentially impacting result reproducibility. Future work addressing this should meticulously document model versions and prompts or consider open-source alternatives. This overall rigorous approach supports our evaluation's reliability, though we acknowledge this specific consideration for LLM-assisted annotation components.

\phead{Internal Validity.}
Due to the high computational cost of fine-tuning and inference evaluation of LLMs, we are unable to conduct experiments under different data splits and randomization states. The randomness in our experiments (e.g., data splitting, explanation annotation, and fine-tuning process) may affect the results. However, we conduct multiple trials to validate the stability of our conclusions for each experimental setup. Although the number of hyperparameter trials for LoRa configuration and the LLM generation method are relatively limited, and due to GPU constraints, we do not attempt to fine-tune larger models, we ensure that these limitations did not fundamentally undermine our primary objective: to explore the potential of conducting vulnerability detection and explanation tasks using LLMs.

\phead{External Validity.}
We chose the advanced open-source code LLM CodeLlama, based on the Llama2 architecture, as our primary model for this research. We test different versions, including 7B and 13B, and evaluate the performance of newer LLM Llama3 and DeepSeek-Coder on vulnerability detection and explanation tasks. However, there are many different architectures of general-purpose open-source LLMs and code models, and our experiments are limited to a few models. Additionally, we conduct experiments using three differently configured open-source project-based vulnerability datasets. Although these datasets include a relatively large amount of vulnerable code, there are still differences compared to code in real development environments. Consequently, our experimental conclusions have an inherent risk of not generalizing to other vulnerability datasets or industrial environments. Nonetheless, our careful selection and thorough evaluation process aim to ensure that our findings remain relevant and insightful within the scope of our study.
\section{Conclusion}
\label{sec:conclusion}

In this paper, we propose \model, a framework designed to detect and explain vulnerabilities using LLMs. The results underscore the potential of LLMs in advancing vulnerability detection and explanation in software security.

Our research provides valuable insights into the fine-tuning of LLMs for vulnerability detection and explanation. It highlights the importance of addressing the data volume bottleneck for training vulnerability LLMs in software development. Furthermore, the quality of annotations is paramount for the success of fine-tuning frameworks. Considering these insights, future work can aim to develop more capable and efficient models that significantly contribute to the field of software security.

\section*{Acknowledgment}
This work was supported by the National Key R\&D Program of China (No. 2024YFB4506400).

\balance
\bibliographystyle{IEEEtran}
\bibliography{paper}

\end{document}